\documentclass[preprint,aps,showpacs]{revtex4}
\usepackage{mathrsfs}
\usepackage{amsmath}
\usepackage{amssymb}
\usepackage{epsfig}
\usepackage{graphicx}
\usepackage{booktabs}
\usepackage{array}
\usepackage{multirow}
\begin{document}
\renewcommand{\arraystretch}{0.5}
\newcommand{\beq}{\begin{eqnarray}}
\newcommand{\eeq}{\end{eqnarray}}
\newcommand{\non}{\nonumber\\ }

\newcommand{\acp}{ {\cal A}_{CP} }
\newcommand{\psl}{ p \hspace{-1.8truemm}/ }
\newcommand{\nsl}{ n \hspace{-2.2truemm}/ }
\newcommand{\vsl}{ v \hspace{-2.2truemm}/ }
\newcommand{\epsl}{\epsilon \hspace{-1.8truemm}/\,  }

\def \cpl{ Chin. Phys. Lett.  }
\def \ctp{ Commun. Theor. Phys.  }
\def \epjc{ Eur. Phys. J. C }
\def \jpg{  J. Phys. G }
\def \npb{  Nucl. Phys. B }
\def \plb{  Phys. Lett. B }
\def \prd{  Phys. Rev. D }
\def \prl{  Phys. Rev. Lett.  }
\def \zpc{  Z. Phys. C }
\def \jhep{ J. High Energy Phys.  }

\title{ Nonleptonic two-body charmless B decays involving a tensor meson in the
Perturbative QCD Approach}
\author{Zhi-Tian Zou}
\author{Xin Yu}
\author{Cai-Dian L\"{u}}\email{lucd@ihep.ac.cn}
\affiliation{Institute  of  High  Energy  Physics  and  Theoretical  Physics Center for Science Facilities,
Chinese Academy of Sciences, Beijing 100049, People's Republic of China }

\date{\today}
\begin{abstract}
Two-body charmless hadronic B decays involving a light tensor meson
in the final states are studied in the perturbative QCD approach
based on $k_T$ factorization. From our calculations, we find that
the decay branching ratios for color allowed tree-dominated decays
$B\to a_{2}^{0}\pi^{+}$ and $B\to a_{2}^{-}\pi^{+}$ modes are of
order $10^{-6}$ and $10^{-5}$, respectively. While other color
suppressed tree-dominated decays have very small branching ratios.
In general, the branching ratios of most  decays are in the range of
$10^{-5}$ to $10^{-8}$, which are   bigger by one or two orders of
magnitude than those predictions obtained in
Isgur-Scora-Grinstein-Wise II model and in the covariant light-front
approach, but consistent with the recent experimental measurements
and the QCD factorization calculations. Since the decays with a
tensor meson emitted from vacuum are prohibited in naive
factorization, the contributions of nonfactorizable and annihilation
diagrams are very important to these decays, which are calculable in
our perturbative QCD approach.
We also give predictions to the direct CP asymmetries, some of which
are large enough for the future experiments to measure. Because we
considered the mixing between $f_{2}$ and $f_{2}'$, the decay rates
are enhanced significantly for some decays involving
$f_{2}^{\prime}$ meson, even with a small mixing angle.

\end{abstract}



\maketitle

\section{Introduction}

In the quark model, all kinds of mesons are classified by the
spin-parity quantum numbers $J^P$. For example, $J^{p}=0^{-}$
denotes pseudoscalar mesons and $J^{p}=2^{+}$ represents tensor
mesons. The p-wave tensor mesons that we study in this paper include
isovector mesons $a_{2}(1320)$, isodoublet states $K_{2}^{*}(1430)$
and two isosinglet mesons $f_{2}(1270)$, $f_{2}^{\prime}(1525)$
\cite{jpg37075021,wwprd83014008}. For these nine tensor mesons, both
orbital angular momentum and the total spin of quarks are equal to
1. Because of the requirement of the Bose statistics of the tensor
meson, the light-cone distribution amplitudes  of tensor mesons are
antisymmetric under the interchange of momentum fractions of the
quark and anti-quark in the flavor SU(3)  limit
\cite{zheng1,zheng2}.

Recently, several experimental measurements about charmless B decay
modes involving a light tensor meson (T) in the final states have
been obtained
\cite{prd82011502,prl97201802,prd79052005,prd78012004,prl96251803,
prd72072003,prd71092003,ifc32229,prl101161801,prd79072006,prd78052005,prd80112001,prd75012006,prd78092008}.
These decays have been studied in the naive factorization approach
\cite{prd491645,prd555581,prd59077504,epjc22683,epjc22695,prd67014002,jpg36095004,arxiv1004.1928,
arxiv1010.3077}, with which it can be easily shown that $\langle
0\mid j^{\mu}\mid T \rangle =\,0$, where $j^{\mu}$ is the $(V\pm A)$
or $(S\pm P)$ current \cite{zheng1,zheng2,epjc22683,epjc22695}. The
factorizable amplitude with a tensor meson emitted vanishes. So
these decays are prohibited in the naive factorization approach. The
branching rations predicted in the naive factorization approach are
too small   compared with the experimental results, which implies
the importance of nonfactorizable and annihilation type
contributions. The recent QCD factorization (QCDF) approach analysis
\cite{zheng2} proved this. It is worth of mentioning that the
perturbative QCD (PQCD) approach \cite{wang7,prd63074009} is almost
the only method   to calculate these kinds of diagrams, without
fitting the experiments.

In this work we shall study charmless $B_{u(d)}\,\rightarrow\,P\,T$
decays in the perturbative QCD approach
 based on the $k_{T}$ factorization. Due to the heavy mass
of B meson, the two light mesons decayed from the B meson are moving
very fast in the rest frame of B meson. The light quarks in the
final state mesons are all collinear; while the light spectator
quark from B meson is soft. Therefore there must be a hard gluon to
kick the light spectator quark in the B meson to form a fast moving
light meson. In this case, the hard process dominates the decay
amplitude, which make it perturbatively calculable. By keeping the
transverse momentum of quarks, the end point singularity in the
collinear factorization can be eliminated. Double logarithm appears
in the QCD radiative corrections due to the additional energy scale
introduced by the transverse momentum. By using the renormalization
group equation, the double logarithm can be resumed and leads to the
Sudakov factor, which effectively suppresses the endpoint
contribution of the distribution amplitude of mesons in the small
momentum region to make the perturbative calculation reliable. The
annihilation diagrams can also  be perturbatively calculated in the
PQCD approach, which is proved to be the dominant strong phase in
$B$ decays for the direct CP asymmetry \cite{laoban}.
Phenomenologically, the PQCD approach has successfully predicted the
direct CP asymmetry in hadronic $B$ decays \cite{laoban} and the
branching ratios of pure annihilation type $B$ decays \cite{cdlv}.

This paper is organized as follows. In Sec.II, we present the
formalism and wave functions of the considered B meson decays. Then
we perform the perturbative calculations for considered decay
channels with the PQCD approach in Sec.III. The numerical results and
phenomenological analysis are given in Sec.IV. Sec.V
contains the main conclusions and a short summary.
Finally, Appendix A contains input parameters and distribution amplitudes used
in this paper and Appendix B gives various functions that enter the
factorization formulae in the PQCD approach.

\section{FORMALISM AND WAVE FUNCTIONs}

The related weak effective Hamiltonian
$H_{eff}$ \cite{rmp68} for charmless $b\rightarrow d(s)$ transitions
can be written as
\begin{eqnarray}
H_{eff}=\frac{G_{F}}{\sqrt{2}}\left\{\sum_{i=1}^{2}\,C_{i}(\mu)
V_{ub}^{*}V_{uD}O_{i}^{u}(\mu)\,-\,V_{tb}^{*}V_{tD}
\sum_{j=3}^{10}\,C_{j}(\mu)O_{j}(\mu)\right\},
\end{eqnarray}
where $V_{ub}$, $V_{uD}$, $V_{tb}$ and $V_{tD}$ are CKM matrix
elements, $D$ denotes the light down quark d or s, and
$C_{i(j)}(\mu)$ are Wilson coefficients at the renormalization scale
$\mu$. $O_{i(j)}(\mu)$ are the well known effective tree (penguin)
operators \cite{rmp68}.

The non-leptonic B meson decays involve  three energy scales,
including the electroweak scale $M_W$, b quark mass scale $M_B$ and
the factorization scale $\sqrt{\overline{\Lambda}M_B}$, where
$\overline{\Lambda} \equiv M_B-m_b$. When the energy scale is higher
than the W boson mass $M_W$, the physics is the electroweak
interaction which can be calculated perturbatively. The physics from
$M_W$ scale to $M_B$ scale is described by the Wilson coefficients
of effective four quark operators, which is the resummation of
leading logarithm by renormalization equations. The physics between
$M_B$ scale and the factorization scale is calculated by the hard
part calculation in the PQCD approach. The physics below the
factorization scale is described by the hadronic wave functions of
mesons, which are nonperturbative but universal for all decay
processes.

In the PQCD approach, the decay amplitude can be factorized into the
convolution of the Wilson coefficients, the hard scattering kernel
and the light-cone wave functions of mesons characterized by
different scales, respectively. Then, for
$B\,\rightarrow\,M_{2}M_{3}$ decays, the decay amplitude is
conceptually written as the convolution,
\begin{eqnarray}
\mathcal
{A}\;\sim\;&&\int\,dx_{1}dx_{2}dx_{3}b_{1}db_{1}b_{2}db_{2}b_{3}db_{3}\nonumber\\
&&\times
Tr[C(t)\Phi_{B}(x_{1},b_{1})\Phi_{M_{2}}(x_{2},b_{2})
\Phi_{M_{3}}(x_{3},b_{3})H(x_{i},b_{i},t)S_{t}(x_{i})e^{-S(t)}],
\end{eqnarray}
where $x_i$ is the longitudinal momentum fractions of valence
quarks, $b_{i}$ is the conjugate space coordinate of the transverse
momentum $k_{iT}$ of the light quarks, and $t$ is the largest scale
in function $H(x_{i},b_{i},t)$. By using the renormalization group
equations, the large logarithms $\ln(m_{W}/t)$ are included in the
Wilson coefficients $C(t)$. By the threshold resummation, the large
double logarithms $\ln^{2}x_{i}$ are summed  to give $S_{t}(x_{i})$
which smears the end-point singularities on $x_{i}$
\cite{prd66094010}. The last term, $e^{-S(t)}$, is the Sudakov
factor which suppresses the soft dynamics effectively
\cite{prd57443}. Thus it makes the perturbative calculation of the
hard part $H$ applicable at intermediate scale, i.e., $m_{B}$ scale.

We will work in the B meson rest frame and employ the light-cone
coordinates for momentum variables. So the B meson momentum is
chosen as $P_{1}=\frac{m_{B}}{\sqrt{2}}(1,\,1,\,\textbf{0}_{T})$.
For the non-leptonic charmless $B\,\rightarrow\,M_{2}M_{3}$ decays,
we assume that the $M_{2}$($M_{3}$) meson moves in the plus(minus) z
direction carrying the momentum $P_{2}$($P_{3}$). Then the momenta
are given by
\begin{eqnarray}
P_{2}=\frac{m_{B}}{\sqrt{2}}(1-r_{3}^{2},\,r_{2}^{2},\,\textbf{0}_{T}),\;P_{3}=\frac{m_{B}}
{\sqrt{2}}(r_{3}^{2},\,1-r_{2}^{2},\,\textbf{0}_{T}),
\end{eqnarray}
where $r_{2}=\frac{m_{M_{2}}}{m_{B}}$ and
$r_{3}=\frac{m_{M_{3}}}{m_{B}}$. The (light-) quark
momenta in B , $M_{2}$ and $M_{3}$ mesons are defined as $k_{1}$,
$k_{2}$ and $k_{3}$, respectively. We choose
\begin{eqnarray}
k_{1}=(x_{1}P_{1}^{+},0,\textbf{k}_{1T}),\;k_{2}=(x_{2}P_{2}^{+},0,\textbf{k}_{2T}),
\;k_{3}=(0,x_{3}P_{3}^{-},\textbf{k}_{3T}).
\end{eqnarray}

For a tensor meson, the polarization tensor
$\epsilon_{\mu\nu}(\lambda)$ with helicity $\lambda$ can be
constructed via the polarization vectors of a vector meson
\cite{zheng1,zheng2}. They are given by
\begin{eqnarray}
\epsilon^{\mu\nu}(\pm2)\,&\equiv&\,\epsilon(\pm1)^{\mu}\epsilon(\pm1)^{\nu},\nonumber\\
\epsilon^{\mu\nu}(\pm1)\,&\equiv&\,\sqrt{\frac{1}{2}}\left[\epsilon(\pm1)^{\mu}\epsilon(0)^{\nu}\,
+\,\epsilon(0)^{\mu}\epsilon(\pm1)^{\nu}\right],\nonumber\\
\epsilon^{\mu\nu}(0)\,&\equiv&\,\sqrt{\frac{1}{6}}\left[\epsilon(+1)^{\mu}\epsilon(-1)^{\nu}\,
+\,\epsilon(-1)^{\mu}\epsilon(+1)^{\nu}\right]\,+\,\sqrt{\frac{2}{3}}\epsilon(0)^{\mu}
\epsilon(0)^{\nu}.
\end{eqnarray}
With the tensor meson moving on the plus direction of the z-axis, the
polarization vectors of the vector meson are chosen as
\begin{eqnarray}
\epsilon^{\mu}(0)=\frac{1}{\sqrt{2}m_{T}}(k_{0}+k_{3},\,k_{0}-k_{3},\,0,\,0),\;
\epsilon^{\mu}(\pm1)=\frac{1}{\sqrt{2}}(0,\,0,\,1,\,\pm i),
\end{eqnarray}
where $k_{0}$ denotes the energy and $k_{3}$ is the magnitude of the
tensor meson momentum in the B meson rest frame. The polarization
tensor satisfies the relations \cite{zheng1,zheng2}
\begin{eqnarray}
\epsilon^{\mu\nu}(\lambda)=\epsilon^{\nu\mu}(\lambda), & \epsilon^{\mu}_{\mu}(\lambda)=0,
 & \nonumber\\
\epsilon^{\mu\nu}(\lambda)P_{\mu}=\epsilon^{\mu\nu}(\lambda)P_{\nu}=0, & \qquad
\epsilon_{\mu\nu}(\lambda)
(\epsilon^{\mu\nu}(\lambda^{\prime}))^{*}=\delta_{\lambda\lambda^{\prime}}. &
\end{eqnarray}
In the following calculation, we define a new
polarization vector $\epsilon_{T}$ for the considered tensor meson for convenience
\cite{wwprd83014008},
\begin{eqnarray}
\epsilon_{T}(\lambda)=\frac{1}{m_{B}}\epsilon_{\mu\nu}(\lambda)P_{B}^{\nu},
\end{eqnarray}
which satisfies
\begin{eqnarray}
\epsilon_{T\mu}(\pm2)=0,
\quad \epsilon_{T\mu}(\pm1)=\frac{\epsilon(0)\cdot P_{B}\epsilon_{\mu}(\pm1)}{\sqrt{2} m_{B}},
\quad \epsilon_{T\mu}(0)=\frac{\sqrt{\frac{2}{3}}\epsilon(0)\cdot P_{B}\epsilon(0)}{m_{B}}.
\end{eqnarray}
One can find that the new vector $\epsilon_{T}$ is similar to the polarization vector $\epsilon$
of a vector meson,
regardless of the related constants
\cite{wwprd83014008}.


In the PQCD approach, we should choose the proper wave functions for
the B meson and light mesons to calculate the decay amplitude. Because the
B meson is a pseudoscalar heavy meson, the two structure
($\gamma_{\mu}\gamma_{5}$) and $\gamma_{5}$ components remain as
leading contributions \cite{wwprd83014008}. Thus the B meson wave function $\Phi_{B}$ is
written as
\begin{eqnarray}
\Phi_{B}=\frac{i}{\sqrt{6}}\left[\left(\makebox[-1.5pt][l]{/}P+m_{B}\right)\gamma_{5}\phi_{B}(x)\right].
\end{eqnarray}
For the distribution amplitude, we can choose
\begin{eqnarray}
\phi_{B}(x,b)=N_{B}x^{2}(1-x)^{2}\exp\left[-\frac{1}{2}\left(\frac{m_{B}x}{\omega_{B}}\right)^{2}
\,-\,\frac{\omega_{B}^{2}b^{2}}{2}\right],
\end{eqnarray}
where $N_{B}$ is the normalization constant.

For the light pseudoscalar meson (P), the wave function is generally
defined as
\begin{eqnarray}
\Phi_{P}(x)\,=\,\frac{i}{\sqrt{6}}\gamma_{5}\left\{\makebox[-1.5pt][l]{/}P\phi_{P}^{A}(x)+m_{0}^{P}
\phi_{P}^{P}(x)
+m_{0}^{P}(\makebox[0pt][l]{/}n\makebox[0pt][l]{/}v-1)\phi_{P}^{T}(x)\right\},
\end{eqnarray}
where $\phi_{P}^{A,P,T}$ and $m_{0}^{P}$ are the distribution
amplitudes and chiral scale parameter of the pseudoscalar
mesons, respectively.  $x$ denotes the momentum fraction carried by
the quark in the meson, and $n=(1,\,0,\,\textbf{0})$ and
$v=(0,\,1,\,\textbf{0})$ are dimensionless light-like unit vectors
pointing to the plus and minus directions, respectively.

The wave functions for a generic tensor meson are defined by
\cite{wwprd83014008}
\begin{eqnarray}
&&\Phi_{T}^{L}\,=\,\frac{1}{\sqrt{6}}\left[m_{T}\makebox[0pt][l]{/}\epsilon_{\bullet
L}^{*}\phi_{T}(x)\,+\,\makebox[0pt][l]{/}\epsilon_{\bullet
L}^{*}\makebox[-1.5pt][l]{/}P\phi_{T}^{t}(x)+m_{T}^{2}\frac{\epsilon_{\bullet}\cdot
v}{P\cdot v}\phi_{T}^{s}(x)\right],\nonumber\\
&&\Phi_{T}^{\perp}\,=\,\frac{1}{\sqrt{6}}\left[m_{T}\makebox[0pt][l]{/}\epsilon_{\bullet
\perp}^{*}\phi_{T}^{v}(x)\,+\,\makebox[0pt][l]{/}\epsilon_{\bullet
\perp}^{*}\makebox[-1.5pt][l]{/}P\phi_{T}^{T}(x)\,+\,m_{T}i\epsilon_{\mu\nu\rho\sigma}
\gamma_{5}\gamma^{\mu}\epsilon_{\bullet
\perp}^{* \nu}n^{\rho}v^{\sigma}\phi_{T}^{a}(x)\right].
\end{eqnarray}
Here $n$ is the moving direction of the tensor meson, and
$v$ is the opposite direction. We adopt the convention
$\epsilon^{0123}=1$. The vector
$\epsilon_{\bullet}\,\equiv\,\frac{\epsilon_{\mu\nu}v^{\nu}}{P\cdot\,
v}m_T$ is related to the polarization tensor. The distribution
amplitudes can be given by \cite{wwprd83014008,zheng1,zheng2}
\begin{eqnarray}
&\phi_{T}(x)\,=\,\frac{f_{T}}{2\sqrt{2N_{c}}}\phi_{\|}(x),\;&\phi_{T}^{t}\,=\,\frac{f_{T}^{\perp}}{2\sqrt{2N_{c}}}h_{\|}^{(t)}(x),
\nonumber\\
&\phi_{T}^{s}(x)\,=\,\frac{f_{T}^{\perp}}{4\sqrt{2N_{c}}}\frac{d}{dx}h_{\|}^{(s)}(x),\;&
\phi_{T}^{T}(x)\,=\,\frac{f_{T}^{\perp}}{2\sqrt{2N_{c}}}\phi_{\perp}(x),\nonumber\\
&\phi_{T}^{v}(x)\,=\,\frac{f_{T}}{2\sqrt{2N_{C}}}g_{\perp}^{(v)}(x),\;&\phi_{T}^{a}(x)\,=\,\frac{f_{T}}{8\sqrt{2N_{c}}}\frac{d}{dx}g_{\perp}^{(a)}(x).
\end{eqnarray}
The asymptotic twist-2 distribution amplitude is given by
\begin{eqnarray}
\phi_{\|,\perp}(x)\,=\,30x(1-x)(2x-1).
\end{eqnarray}
The twist-3 distribution amplitudes are also asymptotic and the
forms are chosen as \cite{wwprd83014008,zheng1,zheng2}
\begin{eqnarray}
&h_{\|}^{(t)}(x)\,=\,\frac{15}{2}(2x-1)(1-6x+6x^{2}),
&h_{\|}^{(s)}(x)\,=\,15x(1-x)(2x-1),\nonumber\\
&g_{\perp}^{(a)}(x)\,=\,20x(1-x)(2x-1),
&g_{\perp}^{(v)}(x)\,=\,5(2x-1)^{3}.
\end{eqnarray}

\section{Perturbative calculation}\label{sec:bcdv}

\begin{figure}[htb]
\begin{center}
\vspace{-1cm} \centerline{\epsfxsize=10 cm \epsffile{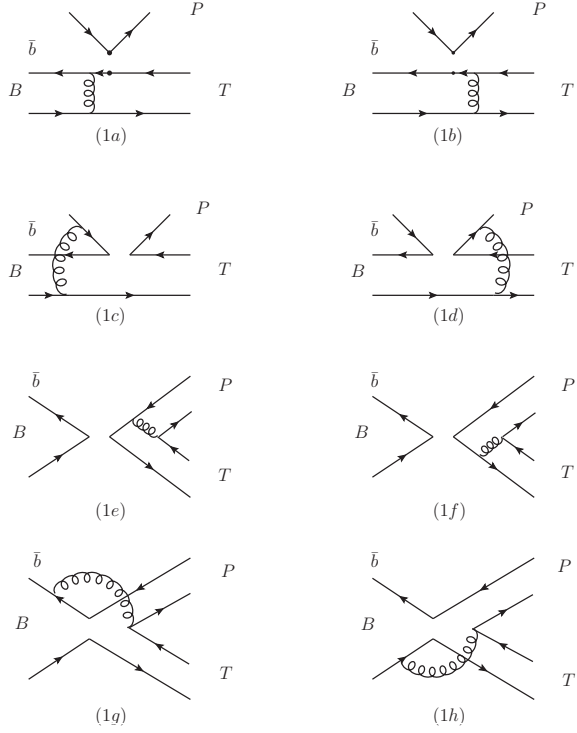}}
\vspace{-3cm} \caption{Diagrams contributing to the
$B\,\rightarrow\,PT$ decays, with a pseudoscalar meson emitted.}
 \label{fig:lodiagram}
 \end{center}
\end{figure}

In this section, we will calculate the hard part $H(t)$, which
includes the effective four quark operators and the necessary hard
gluon connecting the four quark operator with the spectator quark
\cite{lvepjc23275}. There are 8 types of diagrams contributing to
the $B\,\rightarrow\,PT$ decays, shown in Fig.1. From the first two
diagrams of Fig.1, (1a) and (1b), by perturbative QCD calculations,
we gain the decay amplitudes for factorizable emission contribution.
For $(V-A)(V-A)$ current, the amplitude is written as,
\begin{eqnarray}
\mathcal {A}_{eT}^{LL}&=&-8\sqrt{\frac{2}{3}} \pi C_{F} f_{P}
M_{B}^{4}
\int_{0}^{1}\,dx_{1}dx_{3}\,\int_{0}^{\infty}\,b_{1}db_{1}b_{3}db_{3}\,\phi_{B}(x_{1},b_{1})\nonumber\\
&&\times\left\{\left[\phi_{T}(x_{3})(x_{3}+1)-(\phi_{T}^{s}(x_{3})+\phi_{T}^{t}(x_{3}))r_{T}(2x_{3}-1)
\right]h_{ef}(x_{1},x_{3},b_{1},b_{3})E_{ef}(t_{a})\right.\nonumber\\
&&\left.+\left[2r_{T}\phi_{T}^{s}(x_{3})\right]h_{ef}(x_{3},x_{1},b_{3},b_{1})E_{ef}(t_{b})\right\},
\label{ef}
\end{eqnarray}
where  $r_{T}\,=\,\frac{m_{T}}{m_{B}}$, and $C_{F}\,=\,\frac{4}{3}$.
$f_{P}$ is the decay constant of the pseudoscalar meson. The
function $h_{ef}$, $t_{a,b}$ and $E_{ef}$ can be found in Appendix
B. Form Eq.\ref{ef}, we can obtain the $\langle T|V-A|B\rangle$
transition form factor in the PQCD approach.

The operators $O_{5},O_{6},O_{7}$, and $O_{8}$ have the structure of
$(V-A)(V+A)$. In some decay modes, some of these operators will
contribute to the decay amplitude. Because only the axial
part of $(V+A)$ current will contribute to the pseudoscalar meson
production,
we have
\begin{eqnarray}
\mathcal {A}_{eT}^{LR}\,=\,-\mathcal {A}_{eT}^{LL}.
\end{eqnarray}
In some cases, in order to get the right color structure, we must do
a Fierz transformation for these operators. So we obtain
$(S-P)(S+P)$ operators from $(V-A)(V+A)$ ones. The decay amplitude
is,
\begin{eqnarray}
\mathcal {A}_{eT}^{SP}&=&16\sqrt{\frac{2}{3}}C_{F}f_{P}\pi
m_{B}^{4}\int_{0}^{1}\,dx_{1}dx_{3}\int_{0}^{\infty}\,b_{1}db_{1}b_{3}db_{3}\cdot\phi_{B}(x_{1},b_{1})\nonumber\\
&&\times
\left\{\left[\phi_{T}(x_{3})+r_{T}(\phi_{T}^{s}(x_{3})(x_{3}+2)-\phi_{T}^{t}(x_{3})x_{3})\right]r_{0}h_{ef}
(x_{1},x_{3},b_{1},b_{3})E_{ef}(t_{a})\right.\nonumber\\
&&\left.
+\left[2r_{T}r_{0}\phi_{T}^{s}(x_{3})\right]h_{ef}(x_{3},x_{1},b_{3},b_{1})E_{ef}(t_{b})\right\},
\end{eqnarray}
where $r_{0}=m_{0}^{P}/m_{B}$.

For the non-factorizable diagrams Fig.(1c) and (1d), the amplitudes
involve all three wave functions. The integration of $b_{3}$ can be
performed through $\delta$ function $\delta(b_{1}-b_{3})$, leaving
only integration of $b_{1}$ and $b_{2}$. For the (V-A)(V-A),
(V-A)(V+A) and (S-P)(S+P) type operators, the amplitudes are
\begin{eqnarray}
\mathcal {M}_{eT}^{LL}&=&\frac{32}{3}C_{F}\pi
m_{B}^{4}\int_{0}^{1}\,dx_{1}dx_{2}dx_{3}\int_{0}^{\infty}\,b_{1}db_{1}b_{2}db_{2}\,\phi_{B}(x_{1},b_{1})\phi_{P}^{A}(x_{2})\nonumber\\
&&\times
\left\{\left[\phi_{T}(x_{3})(x_{2}-1)+(\phi_{T}^{s}(x_{3})-\phi_{T}^{t}(x_{3}))r_{T}x_{3}\right]\right.\nonumber\\
&&\cdot \left. h_{enf}(x_{1},1-x_{2},x_{3},b_{1},b_{2})E_{enf}(t_{c})\right.\nonumber\\
&&\left.+\left[\phi_{T}(x_{3})(x_{2}+x_{3})-(\phi_{T}^{s}(x_{3})+\phi_{T}^{t}(x_{3}))r_{T}x_{3}\right]\right.\nonumber\\
&& \cdot \left.
h_{enf}(x_{1},x_{2},x_{3},b_{1},b_{2})E_{enf}(t_{d})\right\},\label{eq22}
\end{eqnarray}
\begin{eqnarray}
\mathcal {M}_{eT}^{LR}&=&-\frac{32}{3}C_{F}\pi
r_{0}m_{B}^{4}\int_{0}^{1}\,dx_{1}dx_{2}dx_{3}\int_{0}^{\infty}b_{1}db_{1}b_{2}db_{2}\,\phi_{B}(x_{1},b_{1})\nonumber\\
&&\times\left\{\left[\phi_{P}^{T}(x_{2})(\phi_{T}(x_{3})(x_{2}-1)+r_{T}(\phi_{T}^{t}(x_3)(-x_{2}
+x_{3}+1)+\phi_{T}^{s}(x_{3})(x_{2}+x_{3}-1)))\right.\right.\nonumber\\
&&\left.\left.+\phi_{P}^{P}(\phi_{T}(x_{2}-1)+r_{T}(\phi_{T}^{s}(x_{3})(x_{2}-x_{3}-1)-\phi_{T}^{t}(x_{3})(x_{2}+x_{3}-1)))\right]\right.\nonumber\\
&&\left.\cdot h_{enf}(x_{1},1-x_{2},x_{3},b_{1},b_{2})E_{enf}(t_{c})\right.\nonumber\\
&&\left.+\left[\phi_{P}^{P}(x_{2})(\phi_{T}(x_{3})x_{2}+r_{T}(\phi_{T}^{t}(x_{3})(x_{3}-x_{2})+\phi_{T}^{s}(x_{3})(x_{2}+x_{3})))\right.\right.\nonumber\\
&&\left.\left.+\phi_{P}^{T}(r_{T}(\phi_{T}^{s}(x_{3})(x_{3}-x_{2})+\phi_{T}^{t}(x_{3})(x_{2}+x_{3}))-\phi_{T}(x_{3})x_{2})\right]\right.\nonumber\\
&&\cdot\left.h_{enf}(x_{1},x_{2},x_{3},b_{1},b_{2})E_{enf}(t_{d})\right\},
\end{eqnarray}
\begin{eqnarray}
\mathcal {M}_{eT}^{SP}&=&-\frac{32}{3}C_{F}\pi
m_{B}^{4}\int_{0}^{1}\,dx_{1}dx_{2}dx_{3}\int_{0}^{\infty}\,b_{1}db_{1}b_{2}db_{2}\,\phi_{B}(x_{1},b_{1})\phi_{P}^{A}(x_{2})\nonumber\\
&&\times
\left\{\left[\phi_{T}(x_{3})(x_{2}-x_{3}-1)+(\phi_{T}^{s}(x_{3})+\phi_{T}^{t}(x_{3}))r_{T}x_{3}\right]\right.\nonumber\\
&&\cdot \left. h_{enf}(x_{1},1-x_{2},x_{3},b_{1},b_{2})E_{enf}(t_{c})\right.\nonumber\\
&&\left.+
\left[\phi_{T}(x_{3})x_{2}+(\phi_{T}^{t}-\phi_{T}^{s})r_{T}x_{3}\right]\right.\nonumber\\
&& \cdot \left.
h_{enf}(x_{1},x_{2},x_{3},b_{1},b_{2})E_{enf}(t_{d})\right\}.
\end{eqnarray}

The factorizable annihilation diagrams Fig.(1e) and (1f), the three
kinds of decay amplitudes for these two diagrams are
\begin{eqnarray}
\mathcal {A}_{aT}^{LL}&=&8\sqrt{\frac{2}{3}}C_{F}f_{B}\pi
m_{B}^{4}\int_{0}^{1}\,dx_{2}dx_{3}\int_{0}^{\infty}\,b_{2}db_{2}b_{3}db_{3}\nonumber\\
&&\times\left\{\left[2\phi_{P}^{P}(x_{2})r_{T}r_{0}(\phi_{T}^{s}(x_{3})(x_{3}-2)-\phi_{T}^{t}(x_{3})
x_{3})-\phi_{P}^{A}(x_{2})\phi_{T}(x_{3})(x_{3}-1)\right]\right.\nonumber\\
&&\left.\cdot h_{af}(x_{2},1-x_{3},b_{2},b_{3})E_{af}(t_{e})\right.\nonumber\\
&&\left.+\left[2\phi_{T}^{s}(x_{3})r_{T}r_{0}(\phi_{P}^{T}(x_{2})(x_{2}-1)+\phi_{P}^{P}(x_{2})(x_{2}+1))
-\phi_{P}^{A}(x_{2})\phi_{T}(x_{3})x_{2}\right]\right.\nonumber\\
&&\cdot\left.h_{af}(1-x_{3},x_{2},b_{3},b_{2})E_{af}(t_{f})\right\},
\end{eqnarray}
\begin{eqnarray}
\mathcal {A}_{aT}^{LR}=-\mathcal {A}_{aT}^{LL},
\end{eqnarray}
\begin{eqnarray}
\mathcal {A}_{aT}^{SP}&=&16\sqrt{\frac{2}{3}}C_{F}f_{B}\pi
m_{B}^{4}\int_{0}^{1}\,dx_{2}dx_{3}\int_{0}^{\infty}\,b_{2}db_{2}b_{3}db_{3}\nonumber\\
&&\times\left\{\left[2\phi_{P}^{P}(x_{2})\phi_{T}(x_{3})r_{0}+\phi_{P}^{A}(x_{2})(\phi_{T}^{s}(x_{3})
+\phi_{T}^{t}(x_{3}))r_{T}(x_{3}-1)\right]\right.\nonumber\\
&&\cdot\left.h_{af}(x_{2},1-x_{3},b_{2},b_{3})E_{af}(t_{e})\right.\nonumber\\
&&\left.-\left[x_{2}r_{0}\phi_{T}(x_{3})(\phi_{P}^{T}(x_{2})-\phi_{P}^{P}(x_{2}))+2\phi_{P}^{A}(x_{2})\phi_{T}^{s}(x_{3})r_{T}\right]\right.\nonumber\\
&&\cdot\left.h_{af}(1-x_{3},x_{2},b_{3},b_{2})E_{af}(t_{f})\right\}.
\end{eqnarray}

For the non-factorizable annihilation diagrams Fig.(1g) and (1h),
all three wave functions are involved in the amplitudes. The
integration of $b_{3}$ can be performed by the $\delta$ function
$\delta(b_{2}-b_{3})$. The expressions of contributions of these two
diagrams are
\begin{eqnarray}
\mathcal {M}_{aT}^{LL}&=&\frac{32}{3}C_{F}\pi
m_{B}^{4}\int_{0}^{1}\,dx_{1}dx_{2}dx_{3}\int_{0}^{\infty}\,b_{1}db_{1}b_{2}db_{2}\,\phi_{B}(x_{1},b_{1})\nonumber\\
&&\times \left\{\left[-r_{T}r_{0}\left(\phi_{P}^{T}(x_{2})(\phi_{T}^{s}(x_{3})(x_{2}-1+x_{3})-\phi_{T}^{t}(x_{3})
(x_{2}-1-x_{3}))\right.\right.\right.\nonumber\\
&&\left.\left.\left.+\phi_{P}^{P}(x_{2})(\phi_{T}^{t}(x_{3})(1-x_{2}-x_{3})+\phi_{T}^{s}(x_{3})(x_{2}-x_{3}+3))
\right)+\phi_{P}^{A}(x_{2})\phi_{T}(x_{3})x_{2}\right]\right.\nonumber\\
&&\left.\cdot h_{anf1}(x_{1},x_{2},x_{3},b_{1},b_{2})E_{anf}(t_{g})\right.\nonumber\\
&&\left.+\left[r_{T}r_{0}\left(\phi_{P}^{P}(x_{2})(\phi_{T}^{s}(x_{3})(x_{2}-x_{3}+1)+\phi_{T}^{t}(x_{3})(x_{2}+x_{3}-1))\right.\right.\right.\nonumber\\
&&\left.\left.\left.-\phi_{P}^{T}(x_{2})(\phi_{T}^{t}(x_{3})(x_{2}-x_{3}+1)+\phi_{T}^{s}(x_{3})(x_{2}+x_{3}-1))\right)\right.\right.\nonumber\\
&&\left.\left.+\phi_{P}^{A}(x_{2})\phi_{T}(x_{3})(x_{3}-1)\right]h_{anf2}(x_{1},x_{2},x_{3},b_{1},b_{2})E_{anf}(t_{h})\right\},
\end{eqnarray}
\begin{eqnarray}
\mathcal {M}_{aT}^{LR}&=&\frac{32}{3}C_{F}\pi m_{B}^{4}
\int_{0}^{1}\,dx_{1}dx_{2}dx_{3}\int_{0}^{\infty}b_{1}db_{1}b_{2}db_{2}\,\phi_{B}(x_{1},b_{1})\nonumber\\
&&\times \left\{\left[r_{T}\phi_{P}^{A}(x_{2})(\phi_{T}^{s}(x_{3})-\phi_{T}^{t}(x_{3}))(x_{3}+1)-r_{0}
\phi_{T}(x_{3})(\phi_{P}^{P}(x_{2})+\phi_{P}^{T}(x_{2}))\right.\right.\nonumber\\
&&\cdot\left.\left.(x_{2}-2)\right]h_{anf1}(x_{1},x_{2},x_{3},b_{1},b_{2})E_{anf}(t_{g})\right.\nonumber\\
&&\left.+\left[r_{0}\phi_{T}(x_{3})x_{2}(\phi_{P}^{P}(x_{2})+\phi_{P}^{T}(x_{2}))-r_{T}\phi_{P}^{A}(x_{2})
(\phi_{T}^{s}(x_{3})-\phi_{T}^{t}(x_{3}))(x_{3}-1)\right]\right.\nonumber\\
&&\cdot\left.h_{anf2}(x_{1},x_{2},x_{3},b_{1},b_{2})E_{anf}(t_{h})\right\},
\end{eqnarray}
\begin{eqnarray}
\mathcal {M}_{aT}^{SP}&=&\frac{32}{3}C_{F}\pi
m_{B}^{4}\int_{0}^{1}\,dx_{1}dx_{2}dx_{3}\int_{0}^{\infty}\,b_{1}db_{1}b_{2}db_{2}\,\phi_{B}(x_{1},b_{1})\nonumber\\
&&\times\left\{\left[-r_{T}r_{0}\phi_{P}^{T}(x_{2})(\phi_{T}^{s}(x_{3})(x_{2}-1+x_{3})+\phi_{T}^{t}(x_{3})(x_{2}-1-x_{3}))\right.\right.\nonumber\\
&&\left.\left.+r_{0}r_{T}\phi_{P}^{P}(x_{2})(\phi_{T}^{s}(x_{3})(x_{2}-x_{3}+3)+\phi_{T}^{t}(x_{3})(x_{2}+x_{3}-1))\right.\right.\nonumber\\
&&\left.\left.+\phi_{P}^{A}(x_{2})\phi_{T}(x_{3})(x_{3}-1)\right] h_{anf1}(x_{1},x_{2},x_{3},b_{1},b_{2})E_{anf}(t_{g})\right.\nonumber\\
&&\left.+\left[-r_{0}r_{T}\phi_{P}^{P}(x_{2})(\phi_{T}^{s}(x_{3})(x_{2}+1-x_{3})+\phi_{T}^{t}(x_{3})(1-x_{2}-x_{3}))\right.\right.\nonumber\\
&&\left.\left.-r_{0}r_{T}\phi_{P}^{T}(x_{2})(\phi_{T}^{t}(x_{3})(-x_{2}+x_{3}-1)+\phi_{T}^{s}(x_{3})(x_{2}+x_{3}-1))\right.\right.\nonumber\\
&&\left.\left.+\phi_{P}^{A}(x_{2})\phi_{T}(x_{3})x_{2}\right]h_{anf2}
(x_{1},x_{2},x_{3},b_{1},b_{2})E_{anf}(t_{h})\right\}.\label{eq30}
\end{eqnarray}

\begin{figure}[!htbh]
\begin{center}
\vspace{-1cm} \centerline{\epsfxsize=10 cm \epsffile{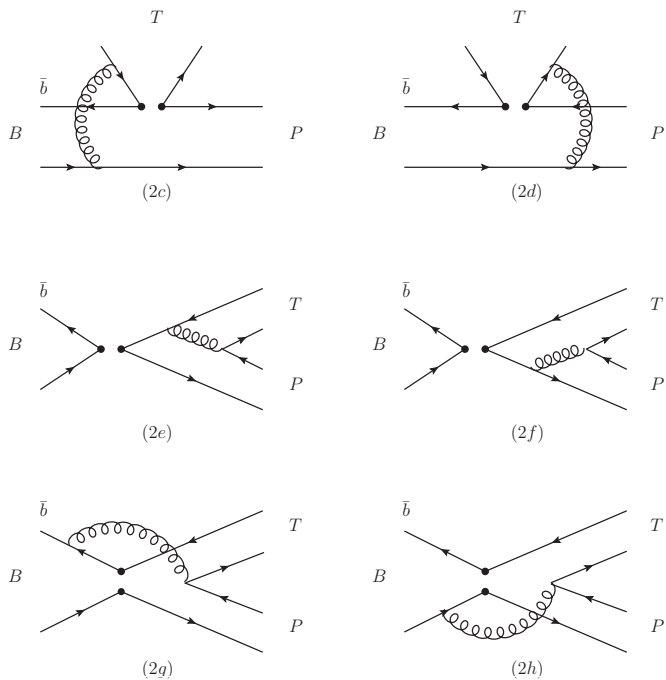}}
\vspace{-3cm} \caption{Diagrams contributing to the
$B\,\rightarrow\,PT$ decays, with a tensor meson emitted.}
\label{fig:lodiagram2}
\end{center}
\end{figure}

If we exchange the pseudoscalar meson and the tensor meson in Fig.1,
the result will be different. Because a tensor meson can not be
produced through (V $\pm$ A) or tensor current, the factorizable
emission diagrams do not contribute to the amplitude of B decays
with a tensor meson emitted \cite{zheng1,zheng2}. Therefore, there
are only six diagrams left shown in Fig.2. The individual decay
amplitudes for these diagrams can be easily deduced from
eq.(\ref{eq22}-\ref{eq30}) by the replacement   of the      wave
functions of the pseudoscalar and the tensor meson,
\begin{eqnarray}
\phi_P^A(x) \rightarrow -\phi_T(x), & \phi_P^P(x) \rightarrow \phi_T^s(x), & \phi_P^T(x) \rightarrow \phi_T^t(x),\nonumber\\
\phi_T(x) \rightarrow -\phi_P^A(x), & \phi_T^s(x) \rightarrow -\phi_P^P(x), & \phi_T^t(x) \rightarrow -\phi_P^T(x),\\
r_T \rightarrow r_0, & r_0 \rightarrow r_T. \nonumber
\end{eqnarray}
In addition, we must add a minus sign to ${M}_{eT}^{SP}$ after applying the above replacement.

For the 39 $B\rightarrow PT$ decay channels, not all the effective
operators contribute to each decay mode. We list the number of
effective operators  contributing  to the individual decay channels
 in the Appendix B for reference.

\section{NUMERICAL RESULTS AND DISCUSSIONS}

For the numerical analysis, we need various input parameters, such
as decay constants, CKM elements and the wave functions, which are
given in Appendix A. The CP-averaged branching ratios  for those
$B\rightarrow PT$ decays with $\Delta S =1$, together with
Isgur-Scora-Grinstein-Wise II (ISGW2) model \cite{prd67014002} and
the QCDF results \cite{zheng2} are shown in table I. The
experimental data are taken from Ref.\cite{jpg37075021} and
Ref.\cite{zheng60}. Similarly, the branching ratios of
$B\,\rightarrow\,PT$ decays with $\Delta S =0$
  calculated in the PQCD approach are shown in Table II.
For illustration, we classify these decays to categories by their
dominant topologies    indicated through the symbols T
(color-allowed tree), C (color-suppressed tree),  P (penguin
emission) and
  PA (penguin annihilation). Although we include also the W annihilation and W exchange diagram
  contributions, none of these channels has dominant contribution
  from these two topology.
For the theoretical uncertainties in our calculation, we estimated
three kinds of them:
 The first errors  are caused by the
uncertainties of the decay constants of tensor mesons; The second
errors are from the decay constant ($f_{B}\,=(\,0.21\,\pm\,0.02)$
GeV) of B meson and the shape parameter
($\omega_{B}\,=\,(0.5\,\pm\,0.05)$ GeV)in the B meson wave function
\cite{zheng1,zheng2,wang7,wang13}; The third errors are estimated
from the unknown next-to-leading order QCD corrections  and  the
power corrections,  characterized by the choice of the
$\lambda_{QCD}\,=\,(0.25\,\pm\,0.05)$ GeV and the variations of the
factorization scales shown in Appendix B, respectively. It is easy
to see that the dominant errors for the branching ratio calculations
are from the non-perturbative wave functions.

It is easy to see that there are large theoretical uncertainties in
any of the individual decay mode calculations. However, we can
reduce the uncertainties by ratios of decay channels. For example,
simple relations among some decay channels are derived in the limit
of SU(3) flavor symmetry
\begin{eqnarray}
&&\mathcal {B}(B^{0}\rightarrow K_{2}^{*0}\pi^{0})\,\sim\,\mathcal
{B}(B^{+}\rightarrow K_{2}^{*+}\pi^{0})\,\sim
\,\frac{1}{2}\,\mathcal {B}(B^{0}\rightarrow
K_{2}^{*+}\pi^{-})\nonumber\\
&&\sim\,\frac{1}{2}\,\mathcal {B}(B^{+}\rightarrow
K_{2}^{*0}\pi^{+}),\nonumber\\
&&\frac{\mathcal {B}(B^{0}\rightarrow a_{2}^{-}K^{+})}{\mathcal
{B}(B^{+}\rightarrow a_{2}^{0}K^{+})}= \frac{\mathcal
{B}(B^{+}\rightarrow a_{2}^{+}K^{0})}{\mathcal {B}(B^{0}\rightarrow
a_{2}^{0}K^{0})}=2.
 \label{guanxi}
\end{eqnarray}
One can find that our results basically agree with the relation
given above within the errors.

\begin{table}
\centering
 \caption{The PQCD predictions of CP-averaged branching
ratios (in units of $10^{-6}$) for $B\rightarrow PT$ decays with
$\Delta S =1$, together with Isgur-Scora-Grinstein-Wise II (ISGW2)
model \cite{prd67014002} and QCDF results \cite{zheng2}. The
experimental data are from Ref.\cite{jpg37075021} and
Ref.\cite{zheng60}.}
\vspace{0.1cm}
\begin{tabular}[t]{l!{\;\;\;\;}c!{\;\;\;}c!{\;\;\;\;\;\;\;}c!{\;\;\;\;\;\;\;}c!{\;\;\;\;\;\;\;}r}
\hline\hline \vspace{0.3cm}
 \multirow{1}{*}{Decay Modes} &\multirow{2}{*}{class}& \multirow{2}{*}{This Work} &\multirow{2}{*}{ISGW2 [24]}&\multirow{2}{*}{QCDF [4]} &\multirow{2}{*}{Expt.}\\
\hline \vspace{0.4cm}
\multirow{1}{*}{$B^{+}\rightarrow K_{2}^{*0}\pi^{+}$}& \multirow{1}{*}{PA} & \multirow{1}{*}{$0.9^{\;+0.2\;+0.2\;+0.3}_{\;-0.2\;-0.2\;-0.2}$} &\multirow{1}{*}{...}&\multirow{1}{*}{$3.1_{-3.1}^{+8.3}$}  &  \multirow{1}{*}{$5.6_{-1.4}^{+2.2}$}\\
\vspace{0.1cm}
$B^{+}\rightarrow K_{2}^{*+}\pi^{0}$ &PA& $0.4^{\;+0.1\;+0.1\;+0.1}_{\;-0.0\;-0.1\;-0.1}$&0.090 &$2.2_{-1.9}^{+4.7}$&...\\
\vspace{0.1cm}
$B^{+}\rightarrow a_{2}^{0}K^{+}$&T,PA&$2.1^{\;+0.7\;+0.6\;+0.6}_{\;-0.6\;-0.5\;-0.5}$&0.31&$4.9_{-4.2}^{+8.4}$&$<45$\\
\vspace{0.1cm}
$B^{+}\rightarrow a_{2}^{+}K^{0}$&PA&$3.1^{\;+0.9\;+0.9\;+1.1}_{\;-0.8\;-0.8\;-0.9}$&0.011&$8.4_{-7.2}^{+16.1}$&...\\
\vspace{0.1cm}
$B^{+}\rightarrow f_{2}K^{+}$&T,PA,P&$11.8^{\;+2.7\;+3.2\;+3.0}_{\;-2.4\;-2.8\;-2.7}$&0.34&$3.8_{-3.0}^{+7.8}$&$1.06_{-0.29}^{+0.28}$\\
\vspace{0.1cm}
$B^{+}\rightarrow f^{\prime}K^{+}$&P,PA&$3.8^{\;+0.4\;+0.9\;+1.0}_{\;-0.4\;-0.8\;-0.8}$&0.004&$4.0_{-3.6}^{+7.4}$&$<7.7$\\
\vspace{0.1cm}
$B^{+}\rightarrow K_{2}^{*+}\eta$&PA,P&$0.8^{\;+0.2\;+0.3\;+0.3}_{\;-0.2\;-0.2\;-0.3}$&0.031&$6.8_{-8.7}^{+13.5}$&$9.1\pm3.0$\\
\vspace{0.1cm}
$B^{+}\rightarrow K_{2}^{*+}\eta^{\prime}$&PA,P&$12.7^{\;+3.7\;+4.5\;+4.0}_{\;-3.2\;-3.5\;-3.5}$&1.41&$12.1_{-12.1}^{+20.7}$&$28.0_{-5.0}^{+5.3}$\\
\vspace{0.1cm}
$B^{0}\rightarrow K_{2}^{*+}\pi^{-}$&PA&$1.0^{\;+0.2\;+0.2\;+0.3}_{\;-0.2\;-0.2\;-0.2}$&...&$3.3_{-3.2}^{+8.5}$&$<6.3$\\
\vspace{0.1cm}
$B^{0}\rightarrow K_{2}^{*0}\pi^{0}$&PA&$0.6^{\;+0.2\;+0.1\;+0.2}_{\;-0.1\;-0.1\;-0.1}$&0.084&$1.2_{-1.3}^{+4.3}$&$<4.0$\\
\vspace{0.1cm}
$B^{0}\rightarrow a_{2}^{-}K^{+}$&T,PA&$5.0^{\;+1.6\;+1.4\;+1.3}_{\;-1.4\;-1.1\;-1.0}$&0.58&$9.7_{-8.1}^{+17.2}$&...\\
\vspace{0.1cm}
$B^{0}\rightarrow a_{2}^{0}K^{0}$&PA&$2.0^{\;+0.5\;+0.4\;+0.6}_{\;-0.5\;-0.4\;-0.5}$&0.005&$4.2_{-3.5}^{+8.3}$&...\\
\vspace{0.1cm}
$B^{0}\rightarrow f_{2}K^{0}$&PA,P&$9.2^{\;+2.0\;+2.5\;+2.6}_{\;-1.8\;-2.1\;-2.2}$&0.005&$3.4_{-3.1}^{+8.5}$&$2.7_{-1.2}^{+1.3}$\\
\vspace{0.1cm}
$B^{0}\rightarrow f_{2}^{\prime}K^{0}$&P,PA&$3.7^{\;+0.3\;+0.7\;+0.9}_{\;-0.4\;-0.8\;-0.9}$&$0.00007$&$3.8_{-3.5}^{+7.3}$&...\\
\vspace{0.1cm}
$B^{0}\rightarrow K_{2}^{*0}\eta$&PA,P&$1.0^{\;+0.2\;+0.3\;+0.3}_{\;-0.2\;-0.2\;-0.3}$&0.029&$6.6_{-8.7}^{+13.5}$&$9.6\pm2.1$\\
\vspace{0.1cm}
$B^{0}\rightarrow K_{2}^{*0}\eta^{\prime}$&PA,P&$11.6^{\;+3.6\;+4.2\;+3.8}_{\;-2.9\;-3.1\;-3.1}$&1.30&$12.4_{-12.4}^{+21.3}$&$13.7_{-3.1}^{+3.2}$\\
 \hline\hline
\end{tabular}\label{S}
\end{table}

\begin{table}[!h]
\centering
 \caption{The PQCD predictions of CP-averaged branching
ratios (in units of $10^{-7}$) for $B\rightarrow PT$ decays with
$\Delta S =0$, together with Isgur-Scora-Grinstein-Wise II (ISGW2)
model \cite{prd67014002} and QCDF results \cite{zheng2}. The
experimental data are from Ref.\cite{jpg37075021} and
Ref.\cite{zheng60}.   }
 \vspace{0.3cm}
\begin{tabular}{l!{\;\;\;\;}c!{\;\;\;}c!{\;\;\;\;\;\;}c!{\;\;\;\;\;}c!{\;\;\;\;\;}r}
\hline\hline \vspace{0.3cm}
\multirow{1}{*}{Decay Modes} &\multirow{1}{*}{class}& \multirow{1}{*}{This Work} &\multirow{1}{*}{ISGW2 [24]}& \multirow{1}{*}{QCDF [4]} &\multirow{1}{*}{Expt.}\\
\hline
 \vspace{0.4cm}
\multirow{1}{*}{$B^{+}\rightarrow a_{2}^{0}\pi^{+}$}&\multirow{1}{*}{T} & \multirow{1}{*}{$29.1^{\;+12.8\;+14.2\;+3.1}_{\;-10.6\;-10.4\;-2.8}$}&\multirow{1}{*}{26.02}& \multirow{1}{*}{$30_{-12}^{+14}$} & \multirow{1}{*}{...}\\
\vspace{0.1cm}
$B^{+}\rightarrow a_{2}^{+}\pi^{0}$&T,C& $0.3^{\;+0.0\;+0.1\;+0.0}_{\;-0.0\;-0.1\;-0.0}$&0.01&$2.4_{-3.1}^{+4.9}$& ...\\
\vspace{0.1cm}
$B^{+}\rightarrow a_{2}^{+}\eta$&C,PA,P &$1.0^{\;+0.3\;+0.4\;+0.4}_{\;-0.3\;-0.3\;-0.3}$&2.94&$1.1_{-1.1}^{+2.8}$&...\\
\vspace{0.1cm}
$B^{+}\rightarrow a_{2}^{+}\eta^{\prime}$&C,PA,P&$3.5^{\;+1.4\;+1.6\;+1.1}_{\;-1.0\;-1.1\;-0.8}$&13.1&$1.1_{-1.2}^{+4.7}$&...\\
\vspace{0.1cm}
$B^{+}\rightarrow f_{2}\pi^{+}$&T&$42.5^{\;+18.9\;+18.9\;+4.2}_{\;-15.4\;-13.9\;-3.9}$&28.74&$27_{-12}^{+14}$&$15.7_{-4.9}^{+6.9}$\\
\vspace{0.1cm}
$B^{+}\rightarrow f_{2}^{\prime}\pi^{+}$&T&$1.2^{\;+0.3\;+0.4\;+0.1}_{\;-0.2\;-0.3\;-0.1}$&0.37&$0.09_{-0.09}^{+0.24}$&...\\
\vspace{0.1cm}
$B^{+}\rightarrow K_{2}^{*+}\bar{K}^{0}$&PA,P&$1.2^{\;+0.2\;+0.2\;+0.3}_{\;-0.2\;-0.2\;-0.3}$&$4.0\times 10^{-4}$&$4.4_{-4.1}^{+7.4}$&...\\
\vspace{0.1cm}
$B^{+}\rightarrow \bar{K}_{2}^{*0}K^{+}$&PA&$0.8^{\;+0.1\;+0.2\;+0.3}_{\;-0.1\;-0.2\;-0.2}$&...&$1.2_{-1.2}^{+5.2}$&...\\
\vspace{0.1cm}
$B^{0}\rightarrow a_{2}^{-}\pi^{+}$&T&$98.9^{\;+35.1\;+42.6\;+5.8}_{\;-29.9\;-32.0\;-9.7}$&48.82&$52_{-18}^{+18}$&$<3000$\\
\vspace{0.1cm}
$B^{0}\rightarrow a_{2}^{+}\pi^{-}$&T,PA&$2.7^{\;+0.5\;+0.8\;+0.4}_{\;-0.3\;-0.5\;-0.3}$&...&$2.1_{-1.7}^{+4.3}$&...\\
\vspace{0.1cm}
$B^{0}\rightarrow a_{2}^{0}\pi^{0}$&C &$4.6^{\;+1.2\;+1.6\;+0.9}_{\;-1.0\;-1.2\;-0.7}$&0.003&$2.4_{-1.9}^{+4.2}$&...\\
\vspace{0.1cm}
$B^{0}\rightarrow a_{2}^{0}\eta$&C,PA,P&$0.6^{\;+0.1\;+0.2\;+0.1}_{\;-0.1\;-0.1\;-0.1}$&1.38&$0.6_{-0.5}^{+1.6}$&...\\
\vspace{0.1cm}
$B^{0}\rightarrow a_{2}^{0}\eta^{\prime}$&C,PA,P&$1.8^{\;+0.6\;+0.7\;+0.4}_{\;-0.5\;-0.6\;-0.4}$&6.15&$0.5_{-0.4}^{+2.2}$&..\\
\vspace{0.1cm}
$B^{0}\rightarrow f_{2}\pi^{0}$&C&$2.8^{\;+0.7\;+0.7\;+0.6}_{\;-0.6\;-0.6\;-0.4}$&0.003&$1.5_{-1.4}^{+4.2}$&...\\
\vspace{0.1cm}
$B^{0}\rightarrow f_{2}^{\prime}\pi^{0}$&P&$0.2^{\;+0.0\;+0.1\;+0.0}_{\;-0.0\;-0.1\;-0.0}$&$4.0\times 10^{-5}$&$0.05_{-0.05}^{+0.12}$&...\\
\vspace{0.1cm}
$B^{0}\rightarrow f_{2}\eta$&C,P,PA&$2.6^{\;+0.7\;+0.8\;+0.7}_{\;-0.5\;-0.6\;-0.6}$&1.52&$1.7_{-1.2}^{+2.3}$&...\\
\vspace{0.1cm}
$B^{0}\rightarrow f_{2}\eta^{\prime}$&C,PA,P&$3.3^{\;+1.0\;+1.1\;+0.9}_{\;-0.8\;-0.9\;-0.9}$&6.8&$1.3_{-1.3}^{+2.2}$&...\\
\vspace{0.1cm}
$B^{0}\rightarrow f_{2}^{\prime}\eta$&PA,P&$0.08^{\;+0.03\;+0.03\;+0.01}_{\;-0.02\;-0.03\;-0.02}$&0.02&$0.02_{-0.03}^{+0.06}$&...\\
\vspace{0.1cm}
$B^{0}\rightarrow f_{2}^{\prime}\eta^{\prime}$&PA,P&$0.09^{\;+0.00\;+0.02\;+0.02}_{\;-0.00\;-0.02\;-0.03}$&0.09&$0.08_{-0.05}^{+0.08}$&...\\
\vspace{0.1cm}
$B^{0}\rightarrow K_{2}^{*+}K^{-}$&PA&$0.16^{\;+0.02\;+0.03\;+0.03}_{\;-0.03\;-0.04\;-0.03}$&...&$0.3_{-0.2}^{+0.7}$&...\\
\vspace{0.1cm}
$B^{0}\rightarrow K_{2}^{*-}K^{+}$&PA&$0.9^{\;+0.1\;+0.3\;+0.2}_{\;-0.1\;-0.1\;-0.2}$&...&$1.3_{-1.0}^{+1.6}$&...\\
\vspace{0.1cm}
$B^{0}\rightarrow K_{2}^{*0}\bar{K}^{0}$&P,PA&$1.5^{\;+0.3\;+0.3\;+0.5}_{\;-0.3\;-0.3\;-0.4}$&$3.0\times 10^{-4}$&$5.4_{-4.9}^{+8.8}$&...\\
\vspace{0.1cm}
$B^{0}\rightarrow \bar{K}_{2}^{*0}K^{0}$&P,PA&$0.8^{\;+0.1\;+0.2\;+0.3}_{\;-0.1\;-0.1\;-0.2}$&...&$2.2_{-2.2}^{+5.4}$&...\\
 \hline\hline
\end{tabular}\label{S2}
\end{table}

Among considered $B\rightarrow PT$ decays, the PQCD predictions for
the CP-averaged branching ratios vary in the range of $10^{-5}$ to
$10^{-8}$. From the numerical results, we  can see that the
predicted branching ratios of penguin-dominated $B\rightarrow PT$
decays in PQCD are larger than those of naive factorization
\cite{prd67014002,jpg36095004,arxiv1010.3077} by one or two orders
of magnitude, but are close to the  QCDF predictions \cite{zheng2}.
For the leading tree-dominated modes such as $a_{2}^{-}\pi^{+}$ and
$f_{2}\pi^{+}$, the predicted results in PQCD are bigger than those
obtained by QCDF \cite{zheng2} but smaller than
Ref.\cite{arxiv1010.3077}. The reason is that the B to tensor form
factor in this work is larger than that used in Ref.\cite{zheng2}.
But for $a_{2}^{0}\pi^{+}$, the result is not larger than  but the
same as Ref.\cite{zheng2}. This is the result of destructive
interference from other topologies. It is worth of remarking that
$B^{0}\,\rightarrow \,K_{2}^{*+}K^{-}$ and $B^{0}\,\rightarrow
\,K_{2}^{*-}K^{+}$ are pure annihilation modes, which can be
perturbatively calculated in the PQCD approach.

The decays with a tensor meson emitted are prohibited in the naive
factorization approach for the reason that a tensor meson can not be
produced from the local (V$\,\pm\,$A) or tensor currents
\cite{zheng1,zheng2}. In order to predict these decay channels, it
is necessary to go beyond the naive factorization framework to
estimate the contributions of the nonfactorizable and annihilation
diagrams. Fortunately, in the PQCD approach, the   contributions of
the nonfactorizable diagrams with a tensor meson emitted (Fig.2, c
and d) are sizable and larger than that of the nonfactorizable
diagrams emitting a pseudoscalar meson (Fig.1, c and d). The reason
is that the asymmetry of the light-cone distribution amplitudes of
the tensor meson makes the contributions from Fig.2(c) and (d)
strengthen with each other, while the situation is contrary for
Fig.1(c) and (d). One can see from Table II that for $B\rightarrow
a_{2}\pi$ decays, the $a_{2}^{+}\pi^{-}$ and $a_{2}^{+}\pi^{0}$
modes are highly suppressed relative to $a_{2}^{-}\pi^{+}$ and
$a_{2}^{0}\pi^{+}$, respectively. It is a natural consequence of
factorization as the tensor meson can not be created from the (V-A)
current. For $B\rightarrow a_{2}^{0}\pi^{+}(a_{2}^{-}\pi^{+})$, the
dominant contribution is from color-allowed factorizable emission
diagrams, while for $B\rightarrow
a_{2}^{+}\pi^{0}(a_{2}^{+}\pi^{-})$, this large contribution is
prohibited for the above reason. Therefore for $B^{+}\rightarrow
a_{2}^{+}\pi^{0}$, the left factorizable emission diagrams are
color-suppressed, and for $B^{0}\rightarrow a_{2}^{+}\pi^{-}$, the
dominant contribution is from nonfactorizable emission diagrams
 suppressed by Wilson coefficient $C_{1}$.

 From table~\ref{tab:suanfu1}, one can see that the factorizable contributions for the  $B^{+}\rightarrow
K_{2}^{*0}\pi^{+}$ and $B^{0}\rightarrow K_{2}^{*+}\pi^{-}$ decays
are 0 because of the emitted meson in these diagrams is the tensor
meson. The     contributions from nonfactorizable diagrams are
suppressed by the small Wilson coefficients $C_{3}$
   and $C_{5}$. Therefore the dominant contribution comes from the penguin annihilation diagrams.
 From table~\ref{S}, one can see that our predictions for the
$B^{+}\rightarrow K_{2}^{*0}\pi^{+}$ and $B^{0}\rightarrow
K_{2}^{*+}\pi^{-}$ decays are much smaller than that from  Ref.
\cite{zheng2}. The reason is that in Ref.\cite{zheng2}, there is an
extremely large contribution from the
   quark loop diagrams. In PQCD approach, the quark loop correction is next-to-leading order and not considered in this
  work.
In the  $B\rightarrow f_{2}K$ decays, we have tree diagram
contribution as well as penguin emission diagram contributions, thus
makes
 the branching ratios much     larger than that of the  $B^{+}\rightarrow
K_{2}^{*0}\pi^{+}$ and $B^{0}\rightarrow K_{2}^{*+}\pi^{-}$ decays.
The current experimental measurements still have very large error
bars.
 We expect the future  experiment to give more information for these decays.

For $B\rightarrow K_{2}^{*}\eta^{(\prime)}$ and $B\rightarrow
a_{2}\eta^{(\prime)}$ decays, one finds that $\mathcal
{B}(B\rightarrow K_{2}^{*}\eta^{\prime})\,\gg\,\mathcal
{B}(B\rightarrow K_{2}^{*} \eta)$ and $\mathcal {B}(B\rightarrow
a_{2}\eta)\,\ll\,\,\mathcal {B}(B\rightarrow a_{2}\eta^{\prime})$.
  For these modes,
both $\eta_{q}$ and $\eta_{s}$ will contribute, but the relative
sign of the $\eta_{s}$ state with respect to the $\eta_{q}$ is
negative for the $\eta$ and positive for the $\eta^{\prime}$, which
leads to a destructive interference between $\eta_{q}$ and
$\eta_{s}$ for $B\rightarrow K_{2}^{*}\eta$ and $B\rightarrow
a_{2}\eta$, but a constructive interference for $B\rightarrow
K_{2}^{*}\eta^{\prime}$ and $B\rightarrow a_{2}\eta^{\prime}$. This
is very similar to the situation for $B\rightarrow K\eta^{(\prime)}$
and $B_{c}\rightarrow K^{+}\eta^{(\prime)}$ decays
\cite{liuxin,liu57}.

\begin{table}[!ht]
\centering
 \caption{The PQCD predictions of  direct CP asymmetries($\%$) for $B\rightarrow PT$ decays with
$\Delta S =1$, comparison with the QCDF results \cite{zheng2}. The
experimental data are from Ref.\cite{jpg37075021}.}
 \vspace{0.3cm}
\begin{tabular}{l!{\;\;\;\;\;\;\;\;\;}c!{\;\;\;\;\;\;\;\;\;}c!{\;\;\;\;\;\;\;\;\;}r}
\hline\hline\vspace{0.3cm}
 \multirow{1}{*}{Decay Modes} & \multirow{1}{*}{This Work} & \multirow{1}{*}{QCDF [4]} &\multirow{1}{*}{Expt.}\\
\hline \vspace{0.4cm}
\multirow{1}{*}{$B^{+}\rightarrow K_{2}^{*0}\pi^{+}$} & \multirow{1}{*}{$-5.5^{\;+0.3\;+2.6\;+1.6}_{\;-0.4\;-0.0\;-1.2}$} & \multirow{1}{*}{$1.6_{-1.8}^{+2.2}$}  & \multirow{1}{*}{$5^{+29}_{-24}$}\\
\vspace{0.1cm}
$B^{+}\rightarrow K_{2}^{*+}\pi^{0}$ & $-6.9^{\;+2.6\;+1.6\;+3.7}_{\;-2.9\;-1.1\;-3.6}$ &$0.2_{-14.8}^{+17.8}$&...\\
\vspace{0.1cm}
$B^{+}\rightarrow a_{2}^{0}K^{+}$&$-52.9^{\;+2.0\;+2.1\;+8.6}_{\;-2.2\;-0.4\;-10.1}$&$27.1_{-35.0}^{+33.3}$&$...$\\
\vspace{0.1cm}
$B^{+}\rightarrow a_{2}^{+}K^{0}$&$2.9^{\;+0.1\;+0.1\;+0.5}_{\;-0.1\;-0.2\;-0.8}$&$-0.6_{-0.8}^{+0.4}$&...\\
\vspace{0.1cm}
$B^{+}\rightarrow f_{2}K^{+}$&$-24.6^{\;+1.5\;+2.4\;+4.6}_{\;-1.0\;-2.6\;-5.9}$&$-39.5_{-25.5}^{+49.4}$&$-68.0_{-17}^{+19}$\\
\vspace{0.1cm}
$B^{+}\rightarrow f^{\prime}K^{+}$&$8.6^{\;+1.5\;+1.4\;+1.5}_{\;-1.6\;-1.0\;-1.8}$&$-0.6_{-6.0}^{+4.3}$&$...$\\
\vspace{0.1cm}
$B^{+}\rightarrow K_{2}^{*+}\eta$&$-5.4^{\;+1.1\;+2.2\;+2.3}_{\;-0.6\;-2.0\;-1.3}$&$1.5_{-5.6}^{+7.4}$&$-45\pm30$\\
\vspace{0.1cm}
$B^{+}\rightarrow K_{2}^{*+}\eta^{\prime}$&$2.0^{\;+0.1\;+0.1\;+0.9}_{\;-0.1\;-0.3\;-0.5}$&$-1.7_{-3.9}^{+3.2}$&$...$\\
\vspace{0.1cm}
$B^{0}\rightarrow K_{2}^{*+}\pi^{-}$&$-17.5^{\;+1.4\;+1.6\;+2.7}_{\;-1.6\;-1.8\;-1.3}$&$1.7_{-5.2}^{+4.2}$&$...$\\
\vspace{0.1cm}
$B^{0}\rightarrow K_{2}^{*0}\pi^{0}$&$-10.7^{\;+0.1\;+1.7\;+1.9}_{\;-0.0\;-1.8\;-1.8}$&$7.1_{-24.1}^{+23.5}$&$...$\\
\vspace{0.1cm}
$B^{0}\rightarrow a_{2}^{-}K^{+}$&$-48.3^{\;+1.9\;+1.3\;+7.1}_{\;-2.4\;-0.3\;-9.9}$&$-21.5_{-35.0}^{+28.9}$&...\\
\vspace{0.1cm}
$B^{0}\rightarrow a_{2}^{0}K^{0}$&$1.9^{\;+0.5\;+0.4\;+0.6}_{\;-0.5\;-0.4\;-0.5}$&$6.7_{-6.9}^{+6.5}$&...\\
\vspace{0.1cm}
$B^{0}\rightarrow f_{2}K^{0}$&$1.2^{\;+0.3\;+0.5\;+0.2}_{\;-0.2\;-0.5\;-0.1}$&$-7.3_{-7.9}^{+8.4}$&$...$\\
\vspace{0.1cm}
$B^{0}\rightarrow f_{2}^{\prime}K^{0}$&$-1.0^{\;+0.1\;+0.0\;+0.0}_{\;-0.3\;-0.1\;-0.1}$&$0.8_{-0.7}^{+1.2}$&...\\
\vspace{0.1cm}
$B^{0}\rightarrow K_{2}^{*0}\eta$&$-5.0^{\;+0.5\;+0.2\;+1.7}_{\;-0.4\;-0.1\;-1.7}$&$3.2_{-4.8}^{+16.5}$&$-7.0\pm19.0$\\
\vspace{0.1cm}
$B^{0}\rightarrow K_{2}^{*0}\eta^{\prime}$&$0.7^{\;+0.1\;+0.1\;+0.3}_{\;-0.0\;-0.0\;-0.2}$&$-2.2_{-4.0}^{+3.3}$&$...$\\
 \hline\hline
\end{tabular}\label{san}
\end{table}

We also give the direct CP asymmetry parameters for those
$B\rightarrow PT$ decays with $\Delta S =1$, together with the QCDF
results \cite{zheng2}   shown in table~\ref{san}. The experimental
data are taken from Ref.\cite{jpg37075021}. Similarly, the direct CP
asymmetry parameters of $B\,\rightarrow\,PT$ decays with $\Delta S
=0$
  calculated in the PQCD approach are shown in Table~\ref{S3}.
The origin of theoretical uncertainties shown in these two tables
are the same as those of the branching ratios in table~\ref{S} and
\ref{S2}. However, the dominant uncertainty here is the third one
from the unknown higher order QCD corrections, since the hadronic
parameter uncertainty mostly cancels due to the fact that the CP
asymmetry is defined as the ratio of branching ratios.

  It is easy
  to see that some channels have very large direct CP asymmetries.
  But many of them have small branching ratios to make them
  difficult for experiments. We recommend the experimenters to
  search for the direct CP asymmetry in the channels like $B^+\to f_2 K^+$, $B^0
  \to a_2^- K^+$, $B^+\to a_2^+\eta'$ and $B^+ \to f_2 \pi^+$, for
  they have both large branching ratios and direct CP asymmetry
  parameters.
  In fact, there are already some experimental measurements for the CP asymmetries
  shown in table~\ref{san} and \ref{S3}. Although the error bars are
  still large, we are happy to see that all these measured entries
  have the same sign as our theoretical calculations. This may imply
  that our approach gives the dominant strong phase in these
  channels.
    The decays $B^{0}(\bar B^{0}) \rightarrow a_{2}^{-}\pi^{+}/
  a_{2}^{+}\pi^{-}$, $B^{0}(\bar B^{0}) \rightarrow K_{2}^{*+}K^{-}/
K_{2}^{*-}K^{+}$ and $B^{0}(\bar B^{0}) \rightarrow K_{2}^{*0} \bar
K^{0}/ \bar K_{2}^{*0}K^{0}$  have a very complicated CP pattern
through the $B^0\bar B^0$ mixing. Four decay amplitudes are involved
for each group of decays with 5 CP parameters to  measure. We refer
the readers to the similar situation for $B^{0}(\bar B^{0})
\rightarrow \rho^{-}\pi^{+}/
  \rho^{+}\pi^{-}$ decays \cite{pirho}.

\begin{table}[!ht]
\centering
 \caption{The PQCD predictions of  direct CP asymmetries($\%$) for $B\rightarrow PT$ decays with
$\Delta S =0$, comparison with the QCDF results \cite{zheng2}. The
experimental data are from Ref.\cite{jpg37075021}.}
 \vspace{0.3cm}
\begin{tabular}{l!{\;\;\;\;\;\;\;\;\;}c!{\;\;\;\;\;\;\;\;\;}c!{\;\;\;\;\;\;\;\;\;}r}
\hline\hline \vspace{0.3cm}
\multirow{1}{*}{Decay Modes} & \multirow{1}{*}{This Work} & \multirow{1}{*}{QCDF [4]} &\multirow{1}{*}{Expt.}\\
\hline \vspace{0.4cm}
\multirow{1}{*}{$B^{+}\rightarrow a_{2}^{0}\pi^{+}$} & \multirow{1}{*}{$-0.6^{\;+0.1\;+0.4\;+0.2}_{\;-0.1\;-0.5\;-0.6}$}& \multirow{1}{*}{$9.6_{-46.6}^{+47.9}$} & \multirow{1}{*}{...}\\
\vspace{0.1cm}
$B^{+}\rightarrow a_{2}^{+}\pi^{0}$ & $-5.8^{\;+0.1\;+21.3\;+75.8}_{\;-0.1\;-12.4\;-44.7}$&$-24.3_{-75.7}^{+124.3}$& ...\\
\vspace{0.1cm}
$B^{+}\rightarrow a_{2}^{+}\eta$    &$-90.9^{\;+8.4\;+9.6\;+12.3}_{\;-3.7\;-1.0\;-5.1}$&$27.6_{-127.6}^{+73.4}$&...\\
\vspace{0.1cm}
$B^{+}\rightarrow a_{2}^{+}\eta^{\prime}$&$-44.5^{\;+0.8\;+1.3\;+6.8}_{\;-0.5\;-0.2\;-8.8}$&$31.3_{-131.3}^{+61.3}$&...\\
\vspace{0.1cm}
$B^{+}\rightarrow f_{2}\pi^{+}$&$27.6^{\;+3.4\;+1.0\;+8.9}_{\;-2.5\;-1.4\;-7.1}$&$60.2_{-72.3}^{+27.1}$&$41\pm30$\\
\vspace{0.1cm}
$B^{+}\rightarrow f_{2}^{\prime}\pi^{+}$&$0.03^{\;+0.1\;+9.6\;+13.8}_{\;-0.1\;-8.9\;-15.8}$&$0.0$&...\\
\vspace{0.1cm}
$B^{+}\rightarrow K_{2}^{*+}\bar{K}^{0}$&$-43.7^{\;+1.3\;+1.8\;+16.4}_{\;-2.0\;-0.5\;-12.4}$&$30.3_{-33.7}^{+51.2}$&...\\
\vspace{0.1cm}
$B^{+}\rightarrow \bar{K}_{2}^{*0}K^{+}$&$49.5^{\;+4.7\;+3.1\;+23.5}_{\;-4.2\;-4.8\;-13.1}$&$-0.26_{-0.27}^{+0.23}$&...\\
\vspace{0.1cm}
$B^{0}\rightarrow a_{2}^{0}\pi^{0}$ &$53.5^{\;+4.7\;+6.9\;+4.2}_{\;-3.8\;-6.9\;-3.5}$&$-86.2_{-26.4}^{+128.9}$&...\\
\vspace{0.1cm}
$B^{0}\rightarrow a_{2}^{0}\eta$    &$-17.7^{\;+17.7\;+11.2\;+21.8}_{\;-15.7\;-22.6\;-24.5}$&$-76.7_{-19.2}^{+100}$&...\\
\vspace{0.1cm}
$B^{0}\rightarrow a_{2}^{0}\eta^{\prime}$&$-59.9^{\;+0.6\;+10.0\;+7.2}_{\;-0.0\;-6.0\;-7.0}$&$-66.0_{-41.1}^{+154}$&..\\
\vspace{0.1cm}
$B^{0}\rightarrow f_{2}\pi^{0}$&$-9.8^{\;+13.9\;+2.8\;+11.8}_{\;-13.2\;-7.5\;-10.8}$&$-37.2_{-85.5}^{+103.8}$&...\\
\vspace{0.1cm}
$B^{0}\rightarrow f_{2}^{\prime}\pi^{0}$&$-0.7^{\;+2.7\;+1.0\;+6.8}_{\;-2.5\;-1.8\;-6.4}$&$0.0$&...\\
\vspace{0.1cm}
$B^{0}\rightarrow f_{2}\eta$&$-42.5^{\;+1.7\;+1.4\;+9.1}_{\;-1.1\;-1.8\;-9.8}$&$69.7_{-102.7}^{+25.7}$&...\\
\vspace{0.1cm}
$B^{0}\rightarrow f_{2}\eta^{\prime}$&$-0.05^{\;+0.1\;+5.0\;+5.3}_{\;-0.6\;-5.1\;-5.3}$&$82.3_{-94.8}^{+22.9}$&...\\
\vspace{0.1cm}
$B^{0}\rightarrow f_{2}^{\prime}\eta$&$70.9^{\;+0.0\;+11.0\;+11.0}_{\;-2.7\;-15.2\;-12.3}$&$0.0$&...\\
\vspace{0.1cm}
$B^{0}\rightarrow f_{2}^{\prime}\eta^{\prime}$&$45.5^{\;+3.2\;+13.5\;+18.5}_{\;-6.8\;-12.1\;-18.8}$&$0.0$&...\\
\hline\hline
\end{tabular}\label{S3}
\end{table}

For the decays involving $f_{2}^{(\prime)}$ in the final states, we
have taken the $f_{2}-f_{2}^{\prime}$ mixing (Eq.(\ref{ffpmix}))
into account, while in Ref.\cite{zheng2}, $f_{2}$ is considered as
an $(u\bar{u}+d\bar{d})/\sqrt{2}$ state and $f_{2}^{\prime}$ a pure
$s\bar{s}$ state. Although the mixing angle is small, the
interference between $f_{2}^q$ and $f_{2}^{s}$ can bring some
remarkable change. For example, the branching ratio of
$B^{+}\rightarrow f_{2}^{\prime}\pi^{+}$ is bigger than the
prediction in Ref.\cite{zheng2}. This can be understood as follows:
Because of the contribution from the color-allowed factorizable
emission diagrams, although suppressed by the mixing angle, the
contribution of $f_{2}^{q}$ term is at the same level with that of
$f_{2}^{s}$ term. Due to the enhancement from $f_{2}^{q}$ term, the
branching ratio becomes larger than the prediction without taking
the mixing into account. The mixing can also bring remarkable change
to direct CP asymmetry. For $B\rightarrow f_{2}^{\prime}\eta^{(\prime)}$, the direct CP
asymmetries are zero \cite{zheng2}\ when $f_{2}$ is a pure
$s\bar{s}$ state. Since the direct CP asymmetry is proportional to
the interference between the tree and penguin contributions
\cite{laoban}, it should be zero indeed because there are no
contributions of penguin operators when $f_{2}^{\prime}$ is a pure
$s\bar{s}$ state. When taking the mixing into account, $f_{2}^{q}$
term can provide penguin contributions, then the direct CP
asymmetries are no longer zero in this work.

For $B\rightarrow f_{2}\,\eta^{(\prime)}$ and
$f_{2}^{\prime}\,\eta^{(\prime)}$ decays, the relevant final state
mesons contain the same components
$\frac{1}{\sqrt{2}}(u\bar{u}+d\bar{d})$ and $s\bar{s}$, therefore
they have the similar branching ratios. The small differences among
their branching ratios mainly come from the different mixing
coefficients, i.e., $\cos\phi$, $\sin\phi$, $\cos\theta$ and
$\sin\theta$ (see Appendix A).

\section{SUMMARY}

We studied the charmless hadronic $B\rightarrow PT$ decays by
employing the PQCD approach based on the $k_{T}$ factorization. In
addition to usual factorization contributions, we also calculated
the non-factorizable and annihilation type diagrams. From our
numerical calculation and phenomenological analysis, we found the
following results:
\begin{itemize}
\item
The factorizable amplitude with a tensor meson emitted vanishes
because a tensor meson cannot be created from the $(V\pm A)$ or
$(S\pm P)$ currents. For these decay modes, the non-factorizable and
annihilation diagrams' contributions are important. For example,
$B^{+}\rightarrow K_{2}^{*0}\pi^{+}$ and $B^{0}\rightarrow
K_{2}^{*+}\pi^{-}$ have sizable branching ratios because of the
contributions of  penguin   annihilation diagrams.
\item
For penguin-dominated $B\rightarrow PT$ decays, because of the
dynamical penguin enhancement, the predicated branching ratios are
larger by one or two orders of magnitude than those predicted in the
naive factorization approach but close to the QCD factorization
predictions in Ref.\cite{zheng2}
\item
For tree-dominated decay modes, the branching ratios predicted by
PQCD are usually very small except for $a_{2}^{0}\pi^{+}$,
$a_{2}^{-}\pi^{+}$ and $f_{2}\pi^{+}$ modes with branching ratios of
order $10^{-6}$ or even larger. This basically agrees with the
situation in Ref.\cite{zheng2} and Ref.\cite{arxiv1010.3077}.
\item
For $B\rightarrow K_{2}^{*}\eta^{(\prime)}$ decays, we find
$\mathcal {B}(B\rightarrow K_{2}^{*}\eta^{\prime})\,\gg\,\mathcal
{B}(B\rightarrow K_{2}^{*}\eta)$. This large difference can be
explained by the destructive and constructive interference between
$\eta_{q}$ and $\eta_{s}$.
\item
From our calculation, we find that the interference
between$f_{2}^{q}$ and $f_{2}^{s}$ can bring some remarkable effects
to some decays involving a $f_{2}^{\prime}$ meson in branching ratio
and direct CP asymmetry.
\item
We predict large direct CP asymmetry for some of the $B\to PT$
decays that accessible for the near future experiments.
\end{itemize}

\section*{Acknowledgment}

 We are very grateful to Xin Liu and
Wei Wang for helpful discussions. This work is partially supported
by
 National
Science Foundation of China under the Grant No.11075168.

\begin{appendix}

\section{Input Parameters And Distribution Amplitudes}

The masses and decay constants of tensor mesons are summarized in
Table V.
\begin{table}[!htbh]
\centering
 \caption{The masses and decay constants of light tensor mesons}
 \vspace{0.3cm}
\begin{tabular}{c!{\;\;\;\;\;\;}c!{\;\;\;\;\;\;}c}
\hline\hline
 Tensor(mass(MeV)) & $f_{T}$(MeV) & $f_{T}^{\perp}$(MeV) \\
\hline
$f_{2}(1270)$&$102\,\pm\,6$&$117\,\pm\,25$\\
$f_{2}^{\prime}(1525)$&$126\,\pm\,12$&$65\,\pm\,12$\\
$a_{2}(1320)$&$107\,\pm\,6$&$105\,\pm\,21$\\
$K_{2}^{*}(1430)$&$118\,\pm\,5$&$77\,\pm\,14$\\
 \hline\hline
\end{tabular}\label{S4}
\end{table}
Other input parameters are
\begin{eqnarray}
&&\Lambda_{\overline{MS}}^{f=4}=0.25,\;m_{b}=4.8,\;f_{\pi}=0.131,\;f_{K}=0.16,\nonumber\\
&&\;m_{0}^{\pi}=1.4,\;m_{0}^{K}=1.6,\;m_{0}^{\eta_{q}}=1.07,\;m_{0}^{\eta_{s}}=1.92.
\end{eqnarray}

We adopt the Wolfenstein parameterization for the CKM matrix,
$A=0.808$, $\lambda=0.2253$, $\bar{\rho}=0.132$ and
$\bar{\eta}=0.341$ \cite{jpg37075021}.

The twist-2(3) pseudoscalar meson distribution amplitude(s)
$\phi_{P}^{A} \,(\phi_{P}^{P},\phi_{P}^{T})$ ($P=\pi,K$) can be
parameterized as \cite{zpc48239,jhep01010},
\begin{eqnarray}
&&\phi_{\pi}^{A}(x)\,=\,\frac{3f_{\pi}}{\sqrt{6}}x(1-x)\left[1\,+\,0.44C_{2}^{3/2}(t)\,+\,0.25C_{4}^{3/2}(t)\right],\\
&&\phi_{\pi}^{P}(x)\,=\,\frac{f_{\pi}}{2\sqrt{6}}\left[1\,+\,0.43C_{2}^{1/2}(t)\,+\,0.09C_{4}^{1/2}(t)\right],\\
&&\phi_{\pi}^{T}(x)\,=\,-\frac{f_{\pi}}{2\sqrt{6}}\left[C_{1}^{1/2}(t)\,+\,0.55C_{3}^{1/2}(t)\right],\\
&&\phi_{K}^{A}(x)\,=\,\frac{3f_{K}}{\sqrt{6}}x(1-x)\left[1\,+\,0.17C_{1}^{3/2}(t)\,+\,0.2C_{2}^{3/2}(t)\right],\\
&&\phi_{K}^{P}(x)\,=\,\frac{f_{K}}{2\sqrt{6}}\left[1\,+\,0.24C_{2}^{1/2}(t)\,-\,0.11C_{4}^{1/2}(t)\right],\\
&&\phi_{K}^{T}(x)\,=\,-\frac{f_{K}}{2\sqrt{6}}\left[C_{1}^{1/2}(t)\,+\,0.35C_{3}^{1/2}(t)\right].
\end{eqnarray}
The Gegenbauer polynomials can be defined by
\begin{eqnarray}
&&C_{1}^{1/2}(t)\,=\,t,\;\;\;\;\;\;C_{1}^{3/2}(t)\,=\,3t,\nonumber\\
&&C_{2}^{1/2}(t)\,=\,\frac{1}{2}(3t^{2}-1),\;\;\;C_{2}^{3/2}(t)\,=\,\frac{3}{2}
(5t^{2}-1),\nonumber\\
&&C_{3}^{1/2}(t)\,=\,\frac{1}{2}t(5t^{2}-3),\nonumber\\
&&C_{4}^{1/2}(t)\,=\,\frac{1}{8}(35t^{4}-30t^{2}+3),\,\;C_{4}^{3/2}(t)\,=\,\frac{15}{8}(21t^{4}-14t^{2}+1),
\end{eqnarray}
where $t\,=\,2x-1$. In the above distribution amplitudes for kaon,
the momentum fraction $x$ is carried by the "s" quark.

For the $\eta\,-\,\eta^{\prime}$ system, we use the quark-flavor basis
\cite{liu48}, with $\eta_{q}$ and $\eta_{s}$ defined by
\begin{eqnarray}
\eta_{q}\,=\,\frac{1}{\sqrt{2}}(u\bar{u}+d\bar{d}),\;\;\;\eta_{s}\,=\,s\bar{s}.
\end{eqnarray}
The physical states $\eta$ and $\eta^{\prime}$ can be given by
\begin{eqnarray}
\left(
\begin{array}{c}\vspace{0.1cm}
\eta\\
\eta^{\prime}
\end{array}
\right) \,=\, \left(\begin{array}{cc}\vspace{0.1cm}
\cos\phi & -\sin\phi\\
\sin\phi & \cos\phi
\end{array}
\right) \left(\begin{array}{c}\vspace{0.1cm}
\eta_{q}\\
\eta_{s}
\end{array}
\right)
\end{eqnarray}
The decay constants are related to $f_{q}$ and $f_{s}$ via the same
mixing matrix,
\begin{eqnarray}
\left(
\begin{array}{cc}
\vspace{0.1cm}
f_{\eta}^{q}& f_{\eta}^{s}\\
 f_{\eta^{\prime}}^{q}& f_{\eta^{\prime}}^{s}
\end{array}
\right) \,=\, \left(\begin{array}{cc}
\vspace{0.1cm}
\cos\phi & -\sin\phi\\
\sin\phi & \cos\phi
\end{array}
\right) \left(\begin{array}{cc} \vspace{0.1cm}
f_{q}& 0\\
0&f_{s}
\end{array}
\right).
\end{eqnarray}
The three input parameters $f_{q}$, $f_{s}$ and $\phi$ have been
extracted from related experiments \cite{liu48,liu49}:
\begin{eqnarray}
f_{q}\,=\,(1.07\,\pm\,0.02)f_{\pi},\;\,f_{s}\,=\,(1.34\,\pm\,0.06)f_{\pi},
\;\,\phi\,=\,39.3^{\circ}\,\pm1.0^{\circ}.
\end{eqnarray}

Like the $\eta\,-\,\eta^{\prime}$ mixing, the isoscalar tensor
states $f_{2}(1270)$ and $f_{2}^{\prime}(1525)$ also have a mixing
and can be given by
\begin{eqnarray}
&&f_{2}\,=\,f_{2}^{q}\cos\theta\,+\, f_{2}^{s}\sin\theta,\nonumber\\
&&f_{2}^{\prime}\,=\,f_{2}^{q}\sin\theta\,-\,f_{2}^{s}\cos\theta,
\label{ffpmix}
\end{eqnarray}
where $f_{2}^{q}\,=\,\frac{1}{\sqrt{2}}(u\bar{u}\,+\,d\bar{d})$,
$f_{2}^{s}\,=\,s\bar{s}$ and the mixing angle
$\theta\,=\,5.8^{\circ}$\cite{zheng3}, $7.8^{\circ}$\cite{jpg27807}
or $(9\,\pm\,1)^{\circ}$ \cite{jpg37075021}.

\section{Amplitude And Related Hard Functions}

For each individual decay channel, various effective operators
contribute to the decay amplitude. We summarize the number of
effective operators contributing to every channel in
Table~\ref{tab:suanfu1} and \ref{tab:suanfu2} for the $\Delta S=1$
and $\Delta S=0$, respectively, with
\begin{eqnarray}
a_{1}=\frac{C_{1}}{3}+C_{2},&&\;a_{2}=C_{1}+\frac{C_{2}}{3},\nonumber\\
a_{j}=C_{j}+\frac{C_{j+1}}{3}\,(j=3,5,7,9),&&\;a_{n}=\frac{C_{n-1}}{3}+C_{n}\,(n=4,6,8,10).
\end{eqnarray}

\begin{table}[!htbh]
\caption{The effective operators contributing to each  decay mode
with $\Delta S=1$} \label{tab:suanfu1}
\begin{tabular}{l|c|c|c|c}
\toprule[2pt]
channels &\multicolumn{2}{c|}{emission}& \multicolumn{2}{c}{annihilation} \\
\hline
  & factorizable & non-factorizable & factorizable &
  non-factorizable\\ \hline
$B^{0}\rightarrow
K_{2}^{*+}\pi^{-}$&--&$C_{1},C_{3},C_{5},C_{7},C_{9}$&$a_{4},a_{6},a_{8},a_{10}$&$C_{3},C_{5},C_{7},C_{9}$\\

$B^{0}\rightarrow
a_{2}^{-}K^{+}$&$a_{1},a_{4},a_{6},a_{8},a_{10}$&$C_{1},C_{3},C_{5},C_{7},C_{9}$&$a_{4},a_{6},a_{8},a_{10}$&$C_{3},C_{5},C_{7},C_{9}$\\
\multirow{2}{*}{$B^{0}\rightarrow
a_{2}^{0}K^{0}$}&\multirow{2}{*}{$a_{4},a_{6},a_{8},a_{10}$}&$C_{2},C_{3},C_{5},C_{7},C_{8},$&
\multirow{2}{*}{$a_{4},a_{6},a_{8},a_{10}$}&\multirow{2}{*}{$C_{3},C_{5},C_{7},C_{9}$}\\
&&$C_{9},C_{10}$&& \\
\multirow{2}{*}{$B^{0}\rightarrow
K_{2}^{*0}\pi^{0}$}&\multirow{2}{*}{$a_{2},a_{7},a_{9}$}&$C_{2},C_{3},C_{5},C_{7},C_{8},$&
\multirow{2}{*}{$a_{4},a_{6},a_{8},a_{10}$}&\multirow{2}{*}{$C_{3},C_{5},C_{7},C_{9}$}\\
&&$C_{9},C_{10}$&&\\
 \multirow{2}{*}{$B^{0}\rightarrow
f_{2}^{q}K^{0}$}&\multirow{2}{*}{$a_{4},a_{6},a_{8},a_{10}$}&$C_{2},C_{3},C_{4},C_{5},C_{6},$&
\multirow{2}{*}{$a_{4},a_{6},a_{8},a_{10}$}&\multirow{2}{*}{$C_{3},C_{5},C_{7},C_{9}$}\\
&&$C_{7},C_{8},C_{9},C_{10}$&&\\
 \multirow{2}{*}{$B^{0}\rightarrow
\eta^{q}K_{2}^{*0}$}&\multirow{2}{*}{$a_{2},a_{3},a_{5},a_{7},a_{9}$}&$C_{2},C_{3},C_{4},C_{5},C_{6},
$&\multirow{2}{*}{$a_{4},a_{6},a_{8},a_{10}$}&\multirow{2}{*}{$C_{3},C_{5},C_{7},C_{9}$}\\
&&$C_{7},C_{8},C_{9},C_{10}$&&\\
 \multirow{2}{*}{$B^{0}\rightarrow
f_{2}^{s}K^{0}$}&\multirow{2}{*}{--}&$C_{3},C_{4},C_{5},C_{6},C_{7},$&\multirow{2}{*}{$a_{4},a_{6},a_{8},a_{10}$}&\multirow{2}{*}{$C_{3},C_{5},C_{7},C_{9}$}\\
&&$C_{8},C_{9},C_{10}$&&\\

 \multirow{2}{*}{$B^{0}\rightarrow
\eta^{s}K_{2}^{*0}$}&$a_{3},a_{4},a_{5},a_{6},$&$C_{3},C_{4},C_{5},C_{6},C_{7},$
&\multirow{2}{*}{$a_{4},a_{6},a_{8},a_{10}$}&\multirow{2}{*}{$C_{3},C_{5},C_{7},C_{9}$}\\
&$a_{7},a_{8},a_{9},a_{10}$&$C_{8},C_{9},C_{10}$&&\\

$B^{+}\rightarrow
K_{2}^{*0}\pi^{+}$&--&$C_{3},C_{5},C_{7},C_{9}$&$a_{1},a_{4},a_{6},a_{8},a_{10}$&
$C_{1},C_{3},C_{5},C_{7},C_{9}$\\

$B^{+}\rightarrow
K^{0}a_{2}^{+}$&$a_{4},a_{6},a_{8},a_{10}$&$C_{3},C_{5},C_{7},C_{9}$&$a_{1},a_{4},a_{6},a_{8},a_{10}$&$C_{1},C_{3},C_{5},C_{7},C_{9}$\\

\multirow{2}{*}{$B^{+}\rightarrow
K_{2}^{*+}\pi^{0}$}&\multirow{2}{*}{$a_{2},a_{7},a_{9}$}&$C_{1},C_{2},C_{3},C_{5},C_{7},$&$a_{1},a_{4},a_{6},a_{8},a_{10}$&$C_{1},C_{3},C_{5},C_{7},C_{9}$\\
&&$C_{8},C_{9},C_{10}$&&\\

\multirow{2}{*}{$B^{+}\rightarrow
K^{+}a_{2}^{0}$}&\multirow{2}{*}{$a_{1},a_{4},a_{6},a_{8},a_{10}$}&$C_{1},C_{2},C_{3},C_{5},C_{7},$&$a_{1},a_{4},a_{6},a_{8},a_{10}$&$C_{1},C_{3},C_{5},C_{7},C_{9}$\\
&&$C_{8},C_{9},C_{10}$&&\\

\multirow{2}{*}{$B^{+}\rightarrow
K^{+}f_{2}^{q}$}&\multirow{2}{*}{$a_{1},a_{4},a_{6},a_{8},a_{10}$}&$C_{1},C_{2},C_{3},C_{4},C_{5},$&$a_{1},a_{4},a_{6},a_{8},a_{10}$&$C_{1},C_{3},C_{5},C_{7},C_{9}$\\
&&$C_{6},C_{7},C_{8},C_{9},C_{10}$&&\\

\multirow{2}{*}{$B^{+}\rightarrow
K_{2}^{*+}\eta^{q}$}&\multirow{2}{*}{$a_{2},a_{3},a_{5},a_{7},a_{9}$}&$C_{1},C_{2},C_{3},C_{4},C_{5},$&$a_{1},a_{4},a_{6},a_{8},a_{10}$&$C_{1},C_{3},C_{5},C_{7},C_{9}$\\
&&$C_{6},C_{7},C_{8},C_{9},C_{10}$&&\\

\multirow{2}{*}{$B^{+}\rightarrow
f_{2}^{s}K^{+}$}&\multirow{2}{*}{--}&$C_{3},C_{4},C_{5},C_{6},$&$a_{1},a_{4},a_{6},a_{8},a_{10}$&$C_{1},C_{3},C_{5},C_{7},C_{9}$\\
&&$C_{7},C_{8},C_{9},C_{10}$&&\\

\multirow{2}{*}{$B^{+}\rightarrow
\eta^{s}K_{2}^{*+}$}&$a_{3},a_{4},a_{5},a_{6},$&$C_{3},C_{4},C_{5},C_{6},$&$a_{1},a_{4},a_{6},a_{8},a_{10}$&$C_{1},C_{3},C_{5},C_{7},C_{9}$\\
&$a_{7},a_{8},a_{9},a_{10}$&$C_{7},C_{8},C_{9},C_{10}$&&\\

 \bottomrule[2pt]
\end{tabular}
\end{table}

\begin{table}[!htbh]
\caption{The effective operators contributing to each  decay mode
with $\Delta S=0$} \label{tab:suanfu2}
\begin{tabular}{l|c|c|c|c}
\toprule[2pt]
&\multicolumn{2}{c|}{emission}& \multicolumn{2}{c}{annihilation} \\
\hline
 channels &  factorizable & non-factorizable & factorizable & non-factorizable\\ \hline
\multirow{2}{*}{$B^{0}\rightarrow
f_{2}^{q}\pi^{0}$}&$a_{2},a_{4},a_{6},a_{7},$&$C_{2},C_{3},C_{4},C_{5},C_{6},$&$a_{2},a_{4},a_{6},a_{7},$&$C_{2},C_{3},C_{5},C_{7},$\\
&$a_{8},a_{9},a_{10}$&$C_{7},C_{8},C_{9},C_{10}$&$a_{8},a_{9},a_{10}$&$C_{8},C_{9},C_{10}$\\
\multirow{2}{*}{$B^{0}\rightarrow
\eta^{q}a_{2}^{0}$}&$a_{2},a_{3},a_{4},a_{5},a_{6},$&$C_{2},C_{3},C_{4},C_{5},C_{6},$&$a_{2},a_{4},a_{6},a_{7},$&$C_{2},C_{3},C_{5},C_{7},$\\
&$a_{7},a_{8},a_{9},a_{10}$&$C_{7},C_{8},C_{9},C_{10}$&$a_{8},a_{9},a_{10}$&$C_{8},C_{9},C_{10}$\\
\multirow{2}{*}{$B^{0}\rightarrow
a_{2}^{-}\pi^{+}$}&\multirow{2}{*}{$a_{1},a_{4},a_{6},a_{8},a_{10}$}&\multirow{2}{*}{$C_{1},
C_{3},C_{5},C_{7},C_{9}$}&$a_{2},a_{3},a_{4},a_{5},a_{6},$&$C_{2},C_{3},C_{4},C_{5},C_{6},$\\
&&&$a_{7},_{8},a_{9},a_{10}$&$C_{7},C_{8},C_{9},C_{10}$\\
\multirow{2}{*}{$B^{0}\rightarrow
\pi^{-}a_{2}^{+}$}&\multirow{2}{*}{--}&\multirow{2}{*}{$C_{1},C_{3},C_{5},C_{7},C_{9}$}&$a_{2},a_{3},a_{4},a_{5},a_{6},$&$C_{2},C_{3},C_{4},C_{5},C_{6},$\\
&&&$a_{7},_{8},a_{9},a_{10}$&$C_{7},C_{8},C_{9},C_{10}$\\
\multirow{2}{*}{$B^{0}\rightarrow
a_{2}^{0}\pi^{0}$}&$a_{2},a_{4},a_{6},a_{7},a_{8},$&$C_{2},C_{3},C_{5},C_{7},C_{8},$&$a_{2},a_{3},a_{4},a_{5},a_{6},$&$C_{2},C_{3},C_{4},C_{5},C_{6},$\\
&$a_{9},a_{10}$&$C_{9},C_{10}$&$ a_{7},_{8},a_{9},a_{10}$&$ C_{7},C_{8},C_{9},C_{10}$\\
$B^{0}\rightarrow
f_{2}^{s}\pi^{0}$&--&$C_{4},C_{6},C_{8},C_{10}$&--&--\\
$B^{0}\rightarrow
\eta^{s}a_{2}^{0}$&$a_{3},a_{5},a_{7},a_{9}$&$C_{4},C_{6},C_{8},C_{10}$&--&--\\
\multirow{2}{*}{$B^{0}\rightarrow
f_{2}^{q}\eta^{q}$}&$a_{2},a_{3},a_{4},a_{5},a_{6},$&$C_{2},C_{3},C_{4},C_{5},C_{6},$&$a_{2},a_{3},a_{4},a_{5},a_{6},$&$C_{2},C_{3},C_{4},C_{5},C_{6},$\\
&$ a_{7},a_{8},a_{9},a_{10}$&$C_{7},C_{8},C_{9},C_{10}$&$ a_{7},a_{8},a_{9},a_{10}$&$ C_{7},C_{8},C_{9},C_{10}$\\
$B^{0}\rightarrow
f_{2}^{s}\eta^{s}$&--&--&$a_{3},a_{5},a_{7},a_{9}$&$C_{4},C_{6},C_{8},C_{10}$\\
$B^{0}\rightarrow
f_{2}^{q}\eta^{s}$&$a_{3},a_{5},a_{7},a_{9}$&$C_{4},C_{6},C_{8},C_{10}$&--&--\\
$B^{0}\rightarrow
f_{2}^{s}\eta^{q}$&--&$C_{4},C_{6},C_{8},C_{10}$&--&--\\
$B^{0}\rightarrow
K_{2}^{*+}K^{-}$&--&--&$a_{2},a_{3},a_{5},a_{7},a_{9}$&$C_{2},C_{4},C_{6},C_{8},C_{10}$\\
$B^{0}\rightarrow
K_{2}^{*-}K^{+}$&--&--&$a_{2},a_{3},a_{5},a_{7},a_{9}$&$C_{2},C_{4},C_{6},C_{8},C_{10}$\\
\multirow{2}{*}{$B^{0}\rightarrow
K_{2}^{*0}\bar{K}^{0}$}&\multirow{2}{*}{$a_{4},a_{6},a_{8},a_{10}$}&\multirow{2}{*}{$C_{3},C_{5},C_{7},C_{9}$}&$a_{3},a_{4},a_{5},a_{6},$&$C_{3},C_{4},C_{5},C_{6},$\\
&&&$ a_{7},a_{8},a_{9},a_{10}$&$C_{7},C_{8},C_{9},C_{10}$\\
{$B^{0}\rightarrow
\bar{K}_{2}^{*0}K^{0}$}&\multirow{2}{*}{--}&\multirow{2}{*}{$C_{3},C_{5},C_{7},C_{9}$}&$a_{3},a_{4},a_{5},a_{6},$&$C_{3},C_{4},C_{5},C_{6},$\\
&&&$ a_{7},a_{8},a_{9},a_{10}$&$C_{7},C_{8},C_{9},C_{10}$\\
\multirow{2}{*}{$B^{+}\rightarrow
a_{2}^{0}\pi^{+}$}&\multirow{2}{*}{$a_{1},a_{4},a_{6},a_{8},a_{10}$}&$C_{1},C_{2},C_{3},C_{5},C_{7},$&\multirow{2}{*}{$a_{1},a_{4},
a_{6},a_{8},a_{10}$}&\multirow{2}{*}{$C_{1},C_{3},C_{5},C_{7},C_{9}$}\\
&&$ C_{8},C_{9},C_{10}$&&\\
\multirow{2}{*}{$B^{+}\rightarrow
a_{2}^{+}\pi^{0}$}&$a_{2},a_{4},a_{6},a_{7},a_{8},$&$C_{1},C_{2},C_{3},C_{5},C_{7},$&\multirow{2}{*}
{$a_{1},a_{4},a_{6},a_{8},a_{10}$}&\multirow{2}{*}{$C_{1},C_{3},C_{5},C_{7},C_{9}$}\\
&$ a_{9},a_{10}$&$ C_{8},C_{9},C_{10}$&&\\
\multirow{2}{*}{$B^{+}\rightarrow
f_{2}^{q}\pi^{+}$}&\multirow{2}{*}{$a_{1},a_{4},a_{6},a_{8},a_{10}$}&$C_{1},C_{2},C_{3},C_{4},C_{5},$&
\multirow{2}{*}{$a_{1},a_{4},a_{6},a_{8},a_{10}$}&\multirow{2}{*}{$C_{1},C_{3},C_{5},C_{7},C_{9}$}\\
&&$ C_{6},C_{7},C_{8},C_{9},C_{10}$&&\\
\multirow{2}{*}{$B^{+}\rightarrow
\eta^{q}a_{2}^{+}$}&$a_{2},a_{3},a_{4},a_{5},a_{6},$&$C_{1},C_{2},C_{3},C_{4},C_{5},$&\multirow{2}{*}
{$a_{1},a_{4},a_{6},a_{8},a_{10}$}&\multirow{2}{*}{$C_{1},C_{3},C_{5},C_{7},C_{9}$}\\
&$ a_{7},a_{8},a_{9},a_{10}$&$ C_{6},C_{7},C_{8},C_{9},C_{10}$&&\\
$B^{+}\rightarrow
a_{2}^{+}\eta^{s}$&$a_{3},a_{5},a_{7},a_{9}$&$C_{4},C_{6},C_{8},C_{10}$&--&--\\
$B^{+}\rightarrow
\pi^{+}f_{2}^{s}$&--&$C_{4},C_{6},C_{8},C_{10}$&--&--\\
$B^{+}\rightarrow
K^{+}\bar{K}_{2}^{*0}$&--&$C_{3},C_{5},C_{7},C_{9}$&$a_{1},a_{4},a_{6},a_{8},a_{10}$&$C_{1},C_{3},C_{5},C_{7},C_{9}$\\
$B^{+}\rightarrow
K_{2}^{*+}\bar{K}^{0}$&$a_{4},a_{6},a_{8},a_{10}$&$C_{3},C_{5},C_{7},C_{9}$&$a_{1},a_{4},a_{6},a_{8},a_{10}$&$C_{1},C_{3},C_{5},C_{7},C_{9}$\\
 \bottomrule[2pt]
\end{tabular}
\end{table}

For factorizable emission diagrams Fig.1. (1a) and (1b), the h function
 is given by
\begin{eqnarray}
h_{ef}(x_{1},x_{3},b_{1},b_{3})\,&=&\,K_{0}(\sqrt{x_{1}x_{3}}m_{B}b_{1})\nonumber\\
&&\times\left\{\theta(b_{1}-b_{3})K_{0}\left(\sqrt{x_{3}}m_{B}b_{1}\right)I_{0}\left(\sqrt{x_{3}}m_{B}b_{3}\right)\right.\nonumber\\
&&\left.+\theta(b_{3}-b_{1})K_{0}\left(\sqrt{x_{3}}m_{B}b_{3}\right)I_{0}\left(\sqrt{x_{3}}m_{B}b_{1}\right)\right\}\nonumber\\
&&\times S_{t}(x_{3}).
\end{eqnarray}
The hard scales
\begin{eqnarray}
&&t_{a}\,=\,\max\{\sqrt{x_{3}}m_{B},\,1/b_{1},\,1/b_{3}\}\nonumber\\
&&t_{b}\,=\,\max\{\sqrt{x_{1}}m_{B},\,1/b_{1},\,1/b_{3}\},
\end{eqnarray}
are the maximum energy scales in each diagrams to cancel the large
logarithmic radiative corrections. The $S_{t}$ re-sums the threshold
logarithms $\ln^{2}x$ in the hard kernels to all orders, which is
given by \cite{prd66094010}
\begin{eqnarray}
S_{t}(x)\,=\,\frac{2^{1+2c}\Gamma(3/2+c)}{\sqrt{\pi}\Gamma(1+c)}[x(1-x)]^{c},
\end{eqnarray}
with $c\,=\,0.3$ in this work. In the nonfactorizable contributions,
the $S_{t}(x)$ provides a very small numerical effect to the
amplitude \cite{plb555}. Therefore, we omit the $S_{t}(x)$ in those
contributions.

The evolution factors $E_{ef}(t_{a})$ and $E_{ef}(t_{b})$ in the
matrix elements (see section III) are given by
\begin{eqnarray}
E_{ef}(t)\,=\,\alpha_{s}(t)\exp[-S_{B}(t)-S_{3}(t)].
\end{eqnarray}
The Sudakov exponents are defined as
\begin{eqnarray}
S_{B}(t)\,=\,s\left(x_{1}\frac{m_{B}}{\sqrt{2}},b_{1}\right)\,+\,\frac{5}{3}\int_{1/b_{1}}^{t}\frac{d\bar{\mu}}{\bar{\mu}}\gamma_{q}(\alpha_{s}(\bar{\mu})),
\end{eqnarray}
\begin{eqnarray}
S_{2}(t)\,=\,s\left(x_{2}\frac{m_{B}}{\sqrt{2}},b_{2}\right)\,+\,s\left((1-x_{2})\frac{m_{B}}{\sqrt{2}},b_{2}\right)
\,+\,2\int_{1/b_{2}}^{t}\frac{d\bar{\mu}}{\bar{\mu}}\gamma_{q}(\alpha_{s}(\bar{\mu})),
\end{eqnarray}
\begin{eqnarray}
S_{3}(t)\,=\,s\left(x_{3}\frac{m_{B}}{\sqrt{2}},b_{3}\right)\,+\,s\left((1-x_{3})\frac{m_{B}}{\sqrt{2}},b_{3}\right)
\,+\,2\int_{1/b_{3}}^{t}\frac{d\bar{\mu}}{\bar{\mu}}\gamma_{q}(\alpha_{s}(\bar{\mu})),
\end{eqnarray}
where the $s(Q,b)$ can be found in the Appendix A in the
Ref.\cite{prd63074009}.

For the other diagrams, the related functions are summarized as
follows:
\begin{eqnarray}
&&t_{c}\,=\,\max\{\sqrt{x_{1}x_{3}}m_{B},\sqrt{|1-x_{1}-x_{2}|x_{3}}m_{B},1/b_{1},1/b_{2}\},\nonumber\\
&&t_{d}\,=\,\max\{\sqrt{x_{1}x_{3}}m_{B},\sqrt{|x_{1}-x_{2}|x_{3}}m_{B},1/b_{1},1/b_{2}\},\\
&&E_{enf}(t)\,=\,\alpha_{s}(t)\cdot\exp[-S_{B}(t)-S_{2}(t)-S_{3}(t)]\mid\,_{b_{1}\,=\,b_{3}},
\end{eqnarray}
\begin{eqnarray}
h_{enf}(x_{1},x_{2},x_{3},b_{1},b_{2})\,&=&\,\left[\theta(b_{2}-b_{1})K_{0}(\sqrt{x_{1}x_{3}}m_{B}b_{2})I_{0}(\sqrt{x_{1}x_{3}}m_{B}b_{1})\right.\nonumber\\
&&\left.+\theta(b_{1}-b_{2})K_{0}(\sqrt{x_{1}x_{3}}m_{B}b_{1})I_{0}(\sqrt{x_{1}x_{3}}m_{B}b_{2})\right]\nonumber\\
&&\cdot \left\{\begin{array}{ll}
\frac{i\pi}{2}H_{0}^{(1)}\left(\sqrt{(x_{2}-x_{1})x_{3}}m_{B}b_{2}\right),& x_{2}-x_{1}>0;\\
K_{0}\left(\sqrt{(x_{1}-x_{2})x_{3}}m_{B}b_{2}\right),&x_{1}-x_{2}>0.
\end{array}\right.
\end{eqnarray}
\begin{eqnarray}
&&t_{e}\,=\,\max\{\sqrt{1-x_{3}}m_{B},1/b_{2},1/b_{3}\},\nonumber\\
&&t_{f}\,=\,\max\{\sqrt{x_{2}}m_{B},1/b_{2},1/b_{3}\},\\
&&E_{af}(t)\,=\,\alpha_{s}(t)\cdot \exp[-S_{2}(t)-S_{3}(t)],
\end{eqnarray}
\begin{eqnarray}
h_{af}(x_{2},x_{3},b_{2},b_{3})\,&=&\,(\frac{i\pi}{2})^{2}H_{0}^{(1)}\left(\sqrt{x_{2}x_{3}}m_{B}b_{2}\right)\nonumber\\
&&\left[\theta(b_{2}-b_{3})H_{0}^{(1)}\left(\sqrt{x_{3}}m_{B}b_{2}\right)J_{0}\left(\sqrt{x_{3}}m_{B}b_{3}\right)\right.\,+\nonumber\\
&&\left.\theta(b_{3}-b_{2})H_{0}^{(1)}\left(\sqrt{x_{3}}m_{B}b_{3}\right)J_{0}\left(\sqrt{x_{3}}m_{B}b_{2}\right)\right]\cdot
S_{t}(x_{3}).
\end{eqnarray}
\begin{eqnarray}
&&t_{g}\,=\,\max\{\sqrt{x_{2}(1-x_{3})}m_{B},\sqrt{1-(1-x_{1}-x_{2})}m_{B},1/b_{1},1/b_{2}\}\nonumber\\
&&t_{h}\,=\,\max\{\sqrt{x_{2}(1-x_{3})}m_{B},\sqrt{|x_{1}-x_{2}|(1-x_{3})}m_{B},1/b_{1},1/b_{2}\},\\
&&E_{anf}\,=\,\alpha_{s}(t)\cdot
\exp[-S_{B}(t)-S_{2}(t)-S_{3}(t)]\mid\,_{b_{2}=b_{3}},
\end{eqnarray}
\begin{eqnarray}
h_{anf1}(x_{1},x_{2},x_{3},b_{1},b_{2})\,&=&\,\frac{i\pi}{2}\left[\theta(b_{1}-b_{2})H_{0}^{(1)}\left(\sqrt{x_{2}
(1-x_{3})}m_{B}b_{1}\right)J_{0}\left(\sqrt{x_{2}(1-x_{3})}m_{B}b_{2}\right)\right.\nonumber\\
&&\left.+\theta(b_{2}-b_{1})H_{0}^{(1)}\left(\sqrt{x_{2}(1-x_{3})}m_{B}b_{2}\right)J_{0}\left(\sqrt{x_{2}(1-x_{3})}m_{B}b_{1}\right)\right]\nonumber\\
&&\times K_{0}\left(\sqrt{1-(1-x_{1}-x_{2})x_{3}}m_{B}b_{1}\right),
\end{eqnarray}
\begin{eqnarray}
h_{anf2}(x_{1},x_{2},x_{3},b_{1},b_{2})\,&=&\,\frac{i\pi}{2}\left[\theta(b_{1}-b_{2})H_{0}^{(1)}\left(\sqrt{x_{2}(1-x_{3})}m_{B}b_{1}\right)J_{0}
\left(\sqrt{x_{2}(1-x_{3})}m_{B}b_{2}\right)\right.\nonumber\\
&&\left.+\theta(b_{2}-b_{1})H_{0}^{(1)}\left(\sqrt{x_{2}(1-x_{3})}m_{B}b_{2}\right)J_{0}\left(\sqrt{x_{2}(1-x_{3})}m_{B}b_{1}\right)\right]\nonumber\\
&&\times \left\{\begin{array}{ll}
\frac{i\pi}{2}H_{0}^{(1)}\left(\sqrt{(x_{2}-x_{1})(1-x_{3})}m_{B}b_{1}\right),&\;\;\;x_{1}-x_{2}<0,\\
K_{0}\left(\sqrt{(x_{1}-x_{2})(1-x_{3})}m_{B}b_{1}\right),&\;\;\;x_{1}-x_{2}>0,
\end{array}\right.
\end{eqnarray}
where $H_{0}^{(1)}(z)\,=\,J_{0}(z)\,+\,iY_{0}(z)$.

\end{appendix}


\begin{thebibliography}{99}
\section*{REFERENCES}
\bibitem{jpg37075021}
K. Nakamura et.al. [Particle Data Group], J Phys. G \textbf{37},
075021 (2010).
\bibitem{wwprd83014008}
Wei Wang, Phys. Rev. D \textbf{83}, 014008 (2011).
\bibitem{zheng1}
H. Y. Cheng, Y. Koike and K. C. Yang, Phys. Rev. D \textbf{82},
054019 (2010) [arXiv:1007.3541 [hep-ph]].
\bibitem{zheng2}
Hai-Yang cheng and Kwei-Chou Yang, Phys. Rev. D \textbf{83}, 034001
(2008) [arXiv:1010.3309 [hep-ph]].
\bibitem{prd82011502}
P. del Amo Sanchez et al.[BARBAR Collaboration], Phys. Rev. D
\textbf{82}, 011502 (2010).
\bibitem{prl97201802}
B. Aubert et al. [BARBAR Collaboration], Phys. Rev. Lett.
\textbf{97}, 201802 (2006).
\bibitem{prd79052005}
B. Aubert et al. [BARBAR Collaboration], Phys. Rev. D \textbf{79},
052005 (2009) [arXiv:0901.3703 [hep-ex]].
\bibitem{prd78012004}
B. Aubert et al. [BARBAR Collaboration], Phys. Rev. D \textbf{78},
012004 (2008) [arXiv:0803.4451 [hep-ex]].
\bibitem{prl96251803}
A. Garmash et al. [BELLE Collaboration],  Phys. Rev. Lett.
\textbf{96}, 251803 (2006) [arXiv:hep-ex/0512066].
\bibitem{prd72072003}
 B. Aubert et al. [BARBAR Collaboration], Phys. Rev. D \textbf{72} , 072003 (2005)
  [Errati,-ibid.D \textbf{74}. 099903 (2006)]
  [arXiv:hep-ex/0507004].
\bibitem{prd71092003}
 A. Garmash et al. [BELLE Collaboration], Phys. Rev. D \textbf{71},
 092003 (2005) [arXiv:hep-ex/0412066].
\bibitem{ifc32229}
Y. Unno [BELLE Collaboration], Nuovo Cimento Soc. Ital. Fis. C
\textbf{32} (2009) 229.
\bibitem{prl101161801}
B. Aubert et al. [BARBAR Collaboration], Phys. Rev. Lett.
\textbf{101}, 161801 (2008) [arXiv:0806.4419 [hep-ex]].
\bibitem{prd79072006}
B. Aubert et al. [BARBAR Collaboration],  Phys. Rev. D \textbf{79},
072006 (2009) [arXiv:0902.2051 [hep-ex]].
\bibitem{prd78052005}
B. Aubert et al. [BARBAR Collaboration], Phys. Rev. D \textbf{78}
052005 (2008) [arXiv:0711.4417 [hep-ex]]
\bibitem{prd80112001}
B. Aubert et al. [BARBAR collaboration], Phys. Rev. D \textbf{80},
112001 (2009) [arXiv:0905.3615 [hep-ex]].
\bibitem{prd75012006}
 A. Garmash et al. [BELLE Collaboration], Phys. Rev. D \textbf{75},
 012006 (2007) [arXiv:hep-ex/0610081].
\bibitem{prd78092008}
B. Aubert et al. [BARBAR Collaboration], Phys. Rev. D \textbf{78},
092008 (2008) [arXiv:0808.3586 [hep-ex]].
\bibitem{prd491645}
A. C. Katoch and R. C. Verma, Phys. Rev. D \textbf{49}, 1645
(1994);\textbf{55}, 7315(E) (1997).
\bibitem{prd555581}
G. L\'{o}pez Castro and J. H. Mu\~{n}oz, Phys. Rev. D \textbf{55},
5581 (1997) [arXiv:hep-ph/9702238].
\bibitem{prd59077504}
J. H. Mu\~{n}oz, A. A. Rojas, and G. L\'{o}pez Castro, Phys. Rev. D
\textbf{59}, 077504 (1999).
\bibitem{epjc22683}
C. S. Kim, B. H. Lim and S. Oh, Eur. Phys. J. C \textbf{22}, 683
(2002) [arXiv:hep-ph/0101292].
\bibitem{epjc22695}
C. S. Kim, B. H. Lim and S. Oh, Eur. Phys. J. C \textbf{22}, 695
(2002) [Erratum-ibid. C \textbf{24}, 665 (2002)]
[arXiv:hep-ph/01080504].
\bibitem{prd67014002}
C. S. Kim, B. H. Lim and S. Oh, Phys. Rev. D \textbf{67}, 014002
(2003) [arXiv:hep-ph/0205263].
\bibitem{jpg36095004}
J. H. Mu\`{n}oz and N. Quintero, J. Phys. G \textbf{36}, 095004
(2009) [arXiv:0903.3701 [hep-ph]].
\bibitem{arxiv1004.1928}
N. Sharma and R. C. Verma, arXiv:1004.1928 [hep-ph].
\bibitem{arxiv1010.3077}
N. Sharma, R. Dhir and R. C. Verma, Phys. Rev. D \textbf{83}, 014007
(2011) [arXiv:1010.3077 [hep-ph]].
\bibitem{wang7}
Y. Y. Keum, H. n. Li and A. I. Sanda, Phys. Lett. B \textbf{504}, 6
(2001) [arXiv:hep-ph/0004004]; Phys. Rev. D \textbf{63}, 054008
(2001) [arXiv:hep-ph/0004173].
\bibitem{prd63074009}
 C. D. L\"{u},
K. Ukai and M. Z. Yang, Phys. Rev. D \textbf{63}, 074009 (2001)
[arXiv:hep-ph/0004213].
\bibitem{laoban}
B. H. Hong and C. D. Lu, Sci. China G \textbf{49}, 357 (2006)
[arXiv:hep-ph/0505020].
\bibitem{cdlv}
C. D. L\"{u}, K. Ukai, Eur. Phys. J. C \textbf{28}, 305 (2003); Y.
Li, C. D. L\"{u}, J. Phys. G \text{29}, 2115 (2003); High Energy
Phys. \& Nucl. Phys. 27, 1062 (2003).
\bibitem{rmp68}
G. Buchalla, A. J. Buras and M. E. Lautenbacher, \textbf{ Rev. Mod.
Phys. 68} (1996) 1125; A.J. Buras, [hep-ph/9806471].
\bibitem{prd66094010}
H. N. Li, Phys. Rev. D \textbf{66}, 094010 (2002)
\bibitem{prd57443}
H. N. Li and B. Tseng, Phys. Rev. D \textbf{57}, 443 (1998)
\bibitem{wang13}
T. Kurimoto, H. n. Li and A. I. Sanda, Phys. Rev. D \textbf{65},
014007 (2002); Z. T. Wei and M. Z. Yang; Nucl. Phys. B \textbf{642},
263 (2002); C. D. L\"{u} and M. Z. Yang, Eur. Phys. J. C
\textbf{28}, 515 (2003).
\bibitem{lvepjc23275}
C. D. L\"{u} and M. Z. Yang, Eur. Phys. J. C \textbf{23}, 275-287
(2002).
\bibitem{zheng60}
D. Asner et al. [Heavy Flavor Averaging Group], arXiv:1010.1589
[hep-ex] and online update at
http://www.slac.stanford.edu/xorg/hfag.
\bibitem{liuxin}
Xin Liu, Zhen-Jun Xiao and Cai-Dian L\"{u}, Phys. Rev. D
\textbf{81}, 014022 (2010) [arXiv:0912.1163 [hep-ph]].
\bibitem{liu57}
Z. J. Xiao, Z. Q. Zhang, X. Liu and L. B. Guo, Phys. Rev. D
\textbf{78}, 114001 (2008)
\bibitem{pirho}  Zhou Rui, Gao
Xiangdong, Cai-Dian Lu,  arXiv:1111.0181 [hep-ph]

\bibitem{zpc48239}
V. M. Braun, I. E. Filyanov, Z. Phys. C \textbf{48}, 239 (1990)
\bibitem{jhep01010}
P. Ball, J. High Energy Phys. \textbf{01}, 010 (1999)
\bibitem{liu48}
Th. Feldmann, P. Kroll and B. Stech, Phys. Rev. D \textbf{58},
114006 (1998).
\bibitem{liu49}
R. Escribano and J. M. Frere, J. High Energy. Phys. \textbf{06}
(2005) 029; J. Schechter, A. Subbaraman and H. Weigel, Phys. Rev. D
\textbf{48}, 339 (1993).
\bibitem{zheng3}
 Hai-Yang Cheng and Robert Shrock, Phys. Rev. D \textbf{84}, 094008
 (2011) [arXiv:hep-ph/1109.3877].
\bibitem{jpg27807}
D. M. Li, H. Yu, and Q. X. Shen, J. Phys. G \textbf{27}, 807 (2001).
\bibitem{plb555}
H.-n. Li and K. Ukai, Phys. Lett. B \textbf{555}, 197 (2003).



\end{thebibliography}
\end{document}